\documentclass[12pt]{article}
\usepackage{amssymb}
\usepackage{graphicx}
\usepackage{amsmath}
\usepackage[all]{xy}

\newcommand{\be}{\begin{equation}}
\newcommand{\ee}{\end{equation}}
\newcommand{\bqa}{\begin{eqnarray}}
\newcommand{\eea}{\end{eqnarray}}
\newcommand{\bqas}{\begin{eqnarray*}}
\newcommand{\eeas}{\end{eqnarray*}}

\newcommand{\non}{\nonumber}

\def\brc{\langle}
\def\ckt{\rangle}
\def\ket#1{|{#1}\rangle}

\def\de{\partial}
\def\D{\mathcal{D}}
\def\Tr{\hbox{\rm Tr}}

\def\Im{\hbox{\rm Im}}

\def\c\Imonst{\hbox {\rm const.}}

\def\diag{\hbox{\rm diag}}
 
\def\Tr{{\rm Tr}}    

\def\adt{{\dot\alpha}}

\def\alp{\alpha}
\def\bet{\beta}

\def\tht{\theta}

\def\diag{\hbox{\rm diag}}

\usepackage{latexsym}

\begin{document}

\thispagestyle{empty}

\begin{flushright}
 IFUP-TH/2007-04, \\
{\tt hep-th/yymmnnn} \\
March, 2007 \\
\end{flushright}
\vspace{3mm}

\begin{center}
{\Large \bf
The Magnetic Monopole     \\ ~~~  \\   Seventy-Five  Years Later    }
\footnote{To appear in a special volume of  Lecture Notes in Physics, Springer, 
in honor of the 65th birthday of Gabriele Veneziano. } 
\\[12mm]
\vspace{5mm}

\normalsize
  { \bf
Kenichi Konishi}
\footnote{e-mail: konishi(at)df.unipi.it}

\vspace{3mm} 

{\it Department of Physics, ``E. Fermi'',  University of Pisa \\
and 
\\
 INFN, Sezione di Pisa, \\
Largo Pontecorvo, 3, Ed. C, 56127 Pisa, Italy
} \\
%
%
%
\vspace{5mm}

{\bf Abstract}  \\[5mm]

{\parbox{13cm}{\hspace{5mm}
Non-Abelian monopoles are present in the fully quantum mechanical low-energy effective action of many solvable supersymmetric theories.  They behave perfectly as pointlike particles carrying non-Abelian dual magnetic charges.  They play a crucial role in confinement and in dynamical  symmetry breaking in these theories.   There is a natural identification of these excitations within the semiclassical approach, which involves the flavor symmetry in an essential manner.   
We review in an introductory fashion the  recent development which has led to  a better understanding of the nature and definition of 
non-Abelian monopoles, as well as of their role in confinement and dynamical symmetry breaking in strongly interacting theories.

}}
\end{center}

\newpage

\section{Introduction}

Three quarters of a century have  passed since the introduction of magnetic monopoles in quantum field theory by Dirac \cite{Dirac}.   Our understanding of the soliton sector of spontaneously broken gauge theories  \cite{TH}  is still largely unsatisfactory.  In particular, the development in our understanding of  {\it non-Abelian }  versions of monopoles \cite{NAmonop}-\cite{Strass}  and vortices \cite{Vtx}  have been very slow, in spite of many articles written on these subjects, and 
 in spite of the important role these topological excitations are likely to play   in various areas of physics. For instance,  they might hold the key to the mystery of  quark confinement in Quantum Chromodynamics (QCD). Their quantum mechanical properties are gradually emerging, however, thanks to  an ever improving grasp on the nonperturbative dynamics  in the context of supersymmetric gauge theories. 
Some of the ingredients  of this development include 
the  Seiberg-Witten solution of ${\cal N}=2$ supersymmetric gauge theories and exact  instanton summations,  better understanding of the properties of (super-) conformal field theories,  exact results on the chiral condensates  and symmetry breaking pattern in a wide class of ${\cal N}=1$ supersymmetric gauge theories, and so on. Also,  many new results on non-Abelian {\it vortices} and {\it domain walls} are now available, which are closely related to the problems concerning the monopoles.

It is  the author's opinion that a serious discussion about  confinement  and non-Abelian monopoles today cannot ignore these basic results  from  supersymmetric gauge theories. 
 This lecture presents a review of what the author believes to be some of the most relevant aspects  of this development,  which should serve as an introduction to this very exciting  area of research.

\section{Color confinement} 

One of the profound unsolved problems in the elementary particle physics today is  quark confinement. 
A popular idea, due to 't Hooft and Mandelstam \cite{TM}   holds that the ground state of QCD (quantum chromodynamics)  is a dual superconductor: the quarks are confined by the chromo-electric vortices, analogous to the magnetic Abrikosov-Nielsen-Olesen vortex in the usual type II  superconductors in solid.   The Lagrangian of QCD  
\be   L=  -  \frac{1}{4}  F_{\mu \nu}^{a}  F^{\mu \nu \, a}  +   {\bar \psi } \,  (i \gamma_{\mu} \,{\cal D}^{\mu} + m ) \,\psi
\label{Lagrangian}\ee
 however describes the dynamics of quarks and gluons, and it is not obvious from  (\ref{Lagrangian})   how magnetic (dual) degrees of freedom appear and how they interact. 
One way to detect such degrees of freedom is 't Hooft's Abelian gauge fixing.  One chooses the gauge so that a given field (perhaps some composite of $F_{\mu \nu}^{a} $)     in the adjoint representation to take an   Abelian form 
\be    X  =   \left(\begin{array}{ccc}\lambda_1 & 0 & 0 \\0 & \lambda_2 & 0 \\0 & 0 & \lambda_3\end{array}\right),\qquad 
\lambda_{1}> \lambda_{2}> \lambda_{3}. 
\ee   
For  a generic gauge-field configuration $A_{\mu}(x)$, however, it is not possible to keep the above diagonal form everywhere in ${\bf R}^{4}$.  Near a singularity  $\lambda_{1}= \lambda_{2}$,   diagonalization of the matrix
\be    X= X|_{\lambda_{1}=\lambda_{2}} +  \left(\begin{array}{cc}  C(x) & 0 \\  0 & 0\end{array}\right)
\ee   
where $C$ is a $2 \times 2$ matrix, for instance,  of the form, 
\be   C=    \tau^{i}  (x - x_{0})^{i},
\ee
by a gauge transformation  $U(x)$,   introduces  a magnetic monopole, $A_{i} \simeq  U(x) \, \de \, U(x)^{-1}$. 

Another possibility is to use the Cho-Faddeev-Niemi decomposition \cite{FN}   of the gauge fields (for $SU(2)$)
\be        A_{\mu}^a= C_{\mu} {\bf n}^a + {\tilde \sigma}(x) ( \de_{\mu} {\bf
n}\times
{\bf n})^a + \rho \, \de_{\mu} {\bf n}^a; \qquad   {\tilde \sigma}(x) = 1 +
\sigma(x),    \label{FadNie} \ee
in terms of  the unit vector field ${\bf n}$  and   the Abelian  gauge
field $C_{\mu}$
which live  on $S^2$  and $S^1$ factors,  respectively, of $SU(2)$, 
 and  a charged "scalar"  field
\be    \phi= \rho(x)+ i \sigma(x).\ee
The Wu-Yang  singular monopole solution \cite{Wuyang},  for instance,  corresponds to 
\be    n^a=   \frac{ x^a }{ r}, \quad   C_{\mu}= \phi=0.  \label{WYmono}  \ee
It is possible that these singularities, regularized  {\it e.g.},  by the zero of
$1-|\phi|^2$, somehow manage to behave as dominant degrees of freedom in the ground  state of QCD. 

Whichever way,  a central question is whether the magnetic monopoles of QCD is of Abelian or non-Abelian type.  The 't Hooft-Mandelstam scenario is essentially Abelian.  By assuming that the relevant infrared degrees of freedom are those which signal the singularities of Abelian gauge fixing,   one tacitly makes a highly nontrivial dynamical assumption.

  In this respect, the $SU(2)$  gauge theory  is  an exception, though.    It is  quite possible that in this particular case   't Hooft's (or related)  Abelian
 gauge fixing procedure  allows us to   ``detect''  the correct  magnetic degrees of freedom,  even if the system does not dynamically Abelianize \footnote{This could explain the mysterious success of the Abelian dominance idea in lattice simulations of the pure $SU(2)$ gauge theory,  even if there are no other indications for dynamical  Abelianization.  The author thanks T. Suzuki for useful discussions. }.    The singularities of the Abelian gauge-fixing would signal the presence of the magnetic degrees of freedom, which correspond  \cite{KonTak} just to the  Wu-Yang  monopoles,  Eq.~(\ref{WYmono}).      As the  Cho-Faddeev-Niemi   $n_{a}(x)$  field     parametrizes  $\pi_{2}(S^{2})\sim \pi_{2}(SU(2)/U(1)) = {\bf Z}$,   the magnetic charge of the Wu-Yang monopoles are  the same, and quantized  in the same way,   as the 't Hooft-Polyakov monopoles of  the Georgi-Glashow model.   In  more general  $SU(N)$ theories  with $N\ge 3$,  however,  one does not expect such a lucky situation. If the system does not dynamically Abelianize  to $U(1)^{N-1}$ effective system at some low-energy scales,  it  would not be appropriately  described by an effective Lagrangian describing the Abelian  monopoles  which signal the singularities of the Abelian gauge fixing\footnote{Vice versa, in a system where Abelianization does take place, as in a class of supersymmetric models mentioned in Section~\ref{sec:abelian}  below,  't Hooft's Abelian gauge fixing should be a perfectly 
 valid tool for extracting and studying the relevant infrared degrees of freedom.  }.

  Actually, there is  no hint that such a dynamical scenario   (dynamical Abelianization)  is realized in Nature.   We must seriously consider the much more subtle possibility that somehow non-Abelian, magnetic degrees of freedom play a  role in the physics of confinement and chiral symmetry breaking.   Are there models in which the low-energy dynamics is known and in which non-Abelian magnetic  degrees of freedom play a  central role?

It  does not seem to be  widely known that not only do such systems exist, but that in a sense this  (occurrence of light  non-Abelian monopoles)  is a  most typical dynamical  phenomenon  in a wide class of supersymmetric gauge systems.    The class of models in question  is  $N=2$ supersymmetric theories with $SU$, $SO$ or $USp$  gauge groups   with  quark hypermultiplets in various representations \cite{SW1}-\cite{CKKM}.  Moreover, the class of models in which one can make reliable  analysis about their low-energy behavior, have increased enormously thanks to a more recent work on certain $N=1$  models  \cite{CDSW}   with scalar multiplets in the adjoint representation.   Again, the appearance of  massless, non-Abelian monopoles in their low-energy effective action  is  a rule, rather than an exception,   in these models. 

Of course,  in the context of superconformal theories there are famous examples of non-Abelian dualities such 
as  the Montonen-Olive  duality  in   $N=4$ supersymmetric  theories  \cite{MO}  or  the Seiberg duality  in the 
$N=1$ supersymmetric models  \cite{Sei}  with nontrivial infrared fixed points.  
 
 These problems (conformal invariance and confinement)  are closely related, as the confinement and dynamical symmetry breaking can  often be seen  as the result  of breaking of  (nontrivial)  conformal invariance near an  infrared-fixed point theory.  
 
   Evidently, supersymmetric theories are trying to tell us something important about the non-Abelian monopoles and confinement. 
In what follows we review briefly the old difficulties associated with the semiclassical concepts of non-Abelian monopoles.
 It will be  argued that  the dual group properties  of non-Abelian monopoles occurring in a system with gauge symmetry breaking
$ G   \,\,\,\longrightarrow    \,\,\, H  $    are  best   defined by setting the low-energy  $H$ system in Higgs phase, so that the dual  system  is in confinement  phase.   The transformation  law of the  monopoles follows from that of  monopole-vortex mixed configurations in the system with a large hierarchy of energy scales, $v_{1}\gg  v_{2}$, 
\be
 G   \,\,\,{\stackrel {v_{1}} {\longrightarrow}} \,\,\, H  \,\,\,{\stackrel {v_{2}} {\longrightarrow}} \,\,\,
 {\bf 1},  \label{hierarchy}
\ee
under an unbroken, exact color-flavor diagonal symmetry $H_{C+F}$  This last symmetry is broken by individual soliton vortex,  so the latter develops continuous moduli.    The transformation law among the regular monopoles, which appear at the endpoint of the vortex,   follows from that of the vortices.  This defines, once rewritten in the dual, magnetic variables,  the dual group  ${\tilde H}$  under which the monopoles transform as a multiplet.




\section {Semiclassical ``non-Abelian mono\-poles'':   difficulties}

\subsection{Abelian monopoles} 
A system in which the gauge symmetry is spontaneously broken
\be
  G   \,\,\,{\stackrel {\brc \phi_{1} \ckt    \ne 0} {\longrightarrow}} \,\,\, H   \label{this}
\ee
where $H$ is some non-Abelian subgroup of $G$,   possesses   a set of regular magnetic monopole
solutions  in the  semi-classical approximation.  They are  natural generalizations  of the Abelian   't Hooft-Polyakov monopoles \cite{TH},
 found originally  in the $G=SO(3)$ theory broken to  $H=U(1)$ by a Higgs mechanism.   
  In that theory, the field content is just the  $SU(2)$ 
 gauge fields and  a scalar field in the adjoint representation of the gauge group;    the energy of a static field configuration
 has an expression 
 \be   E = \int d^{3}x [\, \frac{1}{4}  F_{ij}^{a\, 2} +   \frac{1}{2}  (D_{i}\phi^{a})^{2} +  \frac{\lambda}{8}  ( \phi^{a \, 2}-  F^{2})^{2}\,]
  \label{energyc} \ee
 where 
 \[    F_{ij}^{a} =  \de_{i}\, A_{j} -  \de_{j} \, A_{i}  - g\, \epsilon^{abc} \, A_{i}^{b} \, A_{j}^{c}; 
 \]
 while $D_{i}\phi^{a}$  is a covariant derivative, 
\[  D_{i} \phi^{a}=  \de_{i}\phi^{a}-  g\,  \epsilon^{abc}  A_{i}^{b}  \, \phi^{c}. 
\]
Now the static finite energy solution of the equation of motion must behave asymptotically as
\be    \phi^{a} \to   n^{a}(x) \, F, \qquad  n^{a}(x)^{2} =1,
\ee
where the vector field $n^{a}(x)$  clearly label the winding of the map $S^{2} \to S^{2}$,  the first sphere being the space sphere surrounding the monopole, the second sphere representing the vacuum orientation in the group space.   One possibility is $n^{a}$ has a fixed orientation,   such as $n^{a}(x)=(0,0,1)$ everywhere:  this represents a vacuum.    Another possibility is that  $n^{a}$ makes a nontrivial winding in the group space  as $x_{i}$ goes around the sphere, {\it e.g.}
\[  n^{a}(x) =   (\sin \theta \, \cos m \phi,  \sin \theta \, \sin m \phi,   \cos \theta), \qquad   m=\pm 1, \pm 2, \ldots.
\]
This integer labels the homotopy classes 
\[   \pi_{2}(SU(2)/U(1))\sim  \pi_{2}(S^{2}) \sim  {\mathbf Z}
\]
of the  scalar field configurations.  The gauge fields  must reduce to the pure gauge, 
\[   A_{i}^{a} \to   \frac{1}{g}  \epsilon^{abc} \, n^{b}(x) \, \de_{i}\, n^{c}(x)
\]
in order for the energy to be finite. 

The solution of the equation of motion in the nontrivial sectors  can be found by rewriting    Eq. (\ref{energyc})  as 
\be   E = \int d^{3}x [\, \frac{1}{4}  ( F_{ij}^{a}  -    \epsilon_{ijk} \, D_{k} \phi^{a})^{2}
 +   \frac{1}{2} \epsilon_{ijk}  \, F_{ij}^{a}\,  D_{k}\, \phi^{a} +  \frac{\lambda}{8}  ( \phi^{a \, 2}-  F^{2})^{2}  
  \label{energybis} \ee
The crucial observation is that  while  the first and third terms are semi-positive definite, the second term is a total  derivative, 
\[       \frac{1}{2} \epsilon_{ijk}  \, F_{ij}^{a}\,  D_{k}\, \phi^{a} = \de_{k} \, B_{k}, \qquad B_{k}   =  \frac{1}{2} \epsilon_{ijk}  \, F_{ij}^{a}\, \phi^{a}.\]
We used above a useful  identity   for the derivatives for  gauge invariant products
 \[\de_{k}  \, \Tr ( A\, B\, \ldots) =   \Tr ( D_{k}A\, B\, \ldots) +  \Tr (A\,  D_{k} B\, \ldots) +\ldots.\]
Thus the second term of Eq. (\ref{energybis})  represents $F$  times  the ``magnetic'' charge
\[           \int dv \, \nabla \cdot {\bf B} =  \int dS \cdot  {\bf B} =  4 \, \pi g_{m}, \qquad   {\bf B} \sim \frac{g_{m}}{r^{3}}\, {\bf  r}. 
\]
If $\lambda  =0, $  $|\phi^{a\, 2}| \to  F^{2}$  (BPS limit)  the mass is proportional to the magnetic charge,  
$  4 \,\pi \,  g_{m} \, F =\frac{\brc \phi \ckt }{g}   $,  while the field configuration satisfies the linear  BPS equation 
\[   F_{ij}^{a} -    \epsilon_{ijk} \, D_{k} \phi^{a}=0,  
\]
with an explicit (BPS) solution \cite{TH}
\be   A_{i}^{a} =   \epsilon_{aij}  r_{j} \frac{ 1 - K(r) }{ g\,  r^{2}}, \qquad  K(r)=   \frac{g F r}{\sinh g F r},  
\label{BPSsol1} \ee
\be \phi^{a} =    r^{a}  \, \frac{H(r)}{g \, r^{2}}, \qquad  H(r) =   g F r \coth  g F r - 1.   
\label{BPSsol2}\ee

\subsection{Non-Abelian unbroken group  \label{sec:nonamonopoles}} 
When the ``unbroken'' gauge group is non-Abelian, the asymptotic gauge field can be written as 
\begin{equation}   F_{ij} =  \epsilon_{ijk} B_k = 
\epsilon_{ijk}  \frac{ r_k 
}{      r^3}  ({ \beta} \cdot  {\bf H}),           \end{equation}
in an appropriate gauge, where ${\bf H}$ are the diagonal generators of $H$ in the Cartan subalgebra. 
A straightforward generalization of the Dirac's quantization condition leads to 
\begin{equation}  2 \, {\beta \cdot \alpha} \in  { \bf Z}   \label{naqcond}
\end{equation}
where $\alpha$ are the root vectors of $H$.\footnote{This is most easily seen by  considering $\Tr e^{i g \oint  A_{i} \, dx^{i} }$  along  an infinitesimal closed curve on the surface of a sphere surrounding the monopole. By enlarging the loop and re-closing it at the other side of the sphere,  one ends up with 
\[    e^{i g\, \int  d {\bf S} \cdot  {\bf B}}    =  e^{  4\pi i {\bf \beta}\cdot {\bf H} }.
\]
This should be an identity operator:  commuting the above with nondiagonal generators $E_{\alpha}$    yields   Eq. (\ref{naqcond}).
 }    

The constant vectors $\beta$  (with the number of components equal to the rank of the group $H$)  label possible monopoles. 
It is easy to see that the solution of Eq. (\ref{naqcond})  is  that  $\beta$  is any  of the {\it weight vectors} of a group whose  
nonzero roots are given by 
\be   \alpha^{*} = \frac{\alpha}{\alpha \cdot \alpha}. 
\label{dualg}\ee
This is just a standard group theory  theorem:  Eq. (\ref{naqcond})  can in fact  be rewritten as the well-known relation between a 
weight vector and a root vector of any group,   $2 \, {\beta \cdot \alpha^{*}}/ (\alpha^{*} \cdot \alpha^{*})   \in  { \bf Z}$.  

 The group generated by Eq. (\ref{dualg}) is known as the {\it dual}   (we shall call it GNOW dual below) of  $H$,  let us call ${\tilde H}$.  One is thus led to  a set of semi-classical {\it degenerate}  monopoles, with multiplicity   equal to that of a representation of ${\tilde H}$;  this has led to the so-called  GNOW   conjecture, {\i.e.}, that they form a multiplet of the group ${\tilde H}$,   dual of  $H$ \cite{GNO}-\cite{EW}.   
 For simply-laced groups (with the same length of all nonzero roots) such as $SU(N)$, $SO(2N)$, the dual of $H$ is basically the same group, 
except that the allowed representations tell us that 
\be     U(N) \leftrightarrow  U(N);  \qquad   SO(2N)\leftrightarrow  SO(2N) ,  \ee
while 
\be  SU(N) \leftrightarrow  \frac{SU(N)}{{\mathbf Z}_{N}}; \qquad   SO(2N+1)   \leftrightarrow  USp(2N).   \label{below}
\ee
 There is no difficulty in explicitly constructing these degenerate set of mono\-poles \cite{EW}.   The basic idea is to embed  the 't Hooft-Polyakov monopoles in various broken $SU(2)$ subgroups.  The main results are summarized in Appendix \ref{sec:General}, Appendix \ref{sec:Roots}.  These set of monopoles
 constitute the prime candidates for the members of a multiplet of the dual group ${\tilde H}$.

 There are however well-known difficulties with such an interpretation.  The first concerns the topological obstruction discussed
in \cite{CDyons}-\cite{Baisbis}:  in the presence of the classical monopole background, it is not  possible  to define a globally  well-defined set of generators isomorphic to $H$.  As a consequence, no  ``colored dyons''  exist.  In a simplest case with 
the breaking 
\be  SU(3)  \,\,\,{\stackrel {\brc \phi_{1} \ckt    \ne 0} {\longrightarrow}} \,\,\, SU(2) \times U(1),
\label{simplebr}\ee
   this means  that 
    \be { no\,\, monopoles\,\, with \,\, charges }  \quad   ({\underline 2},  1^{*})  \quad  { exist},    \label{cannot} \ee
   where the asterisk indicates a dual, magnetic charge.  

The second  can be regarded as an infinitesimal version of the same difficulty:   certain bosonic zero modes around the monopole solution, corresponding to $H$ gauge transformations,  are non-normalizable (behaving as $r^{-1/2}$ asymptotically).  Thus the standard procedure of quantization leading to  $H$ multiplets of monopoles,  does not work.    Some progress on the check of GNOW duality along this orthodox line of thought,  has  been reported  nevertheless \cite{DFHK},  in the context of  ${\cal N}=4$  supersymmetric gauge theories.  Their approach,  however, requires the consideration of particular class  of  multi monopole systems,  neutral with respect to the non-Abelian  group  (more precisely, non-Abelian part of)  $H$ only. 

Both of these difficulties concern  the transformation properties of the  monopoles  under the subgroup  $H$, while the   relevant question should be  how they transform under the dual group, ${\tilde H}$.  As field transformation groups, $H$ and ${\tilde H}$  are relatively nonlocal, the latter  should look like a nonlocal transformation group  in the original, electric description.

Another related question concerns the {\it multiplicity}   of the monopoles: 
  Take again  the case  of the system with a breaking pattern, Eq. (\ref{simplebr}).  One might argue that there is only one monopole, as  all the degenerate solutions  are related  by the unbroken {\it gauge}  group $H=SU(2)$.\footnote{This interpretation however encounters  the difficulties mentioned above. Also there are cases in which degenerate monopoles occur, which are not simply related  by the group $H$, see below.  }
Or one might say that there are two monopoles, in the sense that  according to the semiclassical GNO classification they are supposed to belong to a  doublet of the dual $SU(2)$ group.  Or, perhaps,   one should conclude that there are infinitely many, continuously related solutions, as  the two solutions obtained by embedding the 't Hooft solutions in $(1,3)$ and $(2,3)$  subspaces, are clearly part of the continuous set of (moduli of) solutions.  In short,  what is the multiplicity ($ {\cal N}$)  of the monopoles:
\be     {\cal N}  =  1, \quad 2,\quad  {\rm or} \quad   \infty \,\,  ?  \label{question}
\ee
  Formulated perhaps more adequately:     
\begin{center}  
{\tt   What is the dual group? 

 How do the degenerate magnetic  monopoles transform among themselves under the dual group?   
 
Which of the semiclassical aspects of monopoles  survive quantum effects?  
}  
\end{center}

In the attempt to answer these questions,   some general considerations seem to be  unavoidable.
The first is  the fact  since  $H$ and $\tilde H $ groups are  non-Abelian  the  dynamics  of the system should enter the problem  in  essential  way. 
For instance, the  non-Abelian $H$ interactions can become strongly-coupled at low energies and can break itself dynamically. This indeed occurs in pure 
${\cal N}=2$ super Yang-Mills theories  ({\it i.e.}, theories without quark hypermultiplets), where the exact quantum mechanical
  result is known in terms of the Seiberg-Witten curves \cite{SW1}-\cite{curves}: see below.  Consider for instance, a pure ${\cal N}=2$,   $SU(N+1)$ gauge theory.  Even though partial breaking, {\it e.g.},  $SU(N+1) \to SU(N) \times U(1)$ looks perfectly possible semi-classically, in an appropriate region of classical degenerate vacua, no such vacua exist quantum mechanically.  In {\it all}   vacua the light monopoles are abelian, the effective, magnetic  gauge group  being    $U(1)^{N}$.

  Generally speaking, the concept of a dual group multiplet  is well-defined  when ${\tilde H}$ interactions are weak (or at worst, conformal).   This however means that one must study the original, electric theory in the regime of  strong coupling, which would usually  make the task  exceedingly difficult.  Fortunately,    in ${\cal N}=2$ supersymmetric gauge theories,  exact Seiberg-Witten  curves  describe the fully quantum mechanical consequences of the strong-interaction dynamics in terms of  weakly-coupled dual magnetic variables.   This is how  we know that the non-Abelian monopoles exist in fully quantum theories \cite{BK}:  in the so-called $r$-vacua of 
softly broken ${\cal N}=2$, $SU(N)$ gauge theory, the light monopoles appear as the dominant  infrared degrees of freedom and  interact  as  pointlike particles having the charges of a fundamental multiplet  ${\underline r}$ of an  effective, dual $SU(r)$  gauge group.
  In an $SU(3)$  gauge theory broken to $SU(2)\times U(1)$ as in (\ref{simplebr}),  with an appropriate number of quark multiplets  ($N_{f}  \ge 4$),  for instance,   light magnetic monopoles carrying  the  charges 
   \be  ({\underline {2^{*}}}, 1^{*}) \ee
  under the dual $SU(2) \times U(1)$
appear  in the low-energy effective action.      (Dual) colored dyons do exist! 
  The distinction between $H$ and ${\tilde H}$ is crucial    ({\it cfr.} Eq. (\ref{cannot})).

  In ${\cal N}=2$, $SU(N)$   SQCD   with  $N_{f}$ flavors,  light non-Abelian monopoles with $SU(r)$  dual gauge group appear  for   $r \le \frac{N_{f}}{2}$ only.  Such a limit clearly reflects the dynamics of the soliton monopoles under  renormalization group:  the effective low-energy  gauge group must be either infrared free or  conformally invariant, in order  for  the monopoles to emerge as  recognizable low-energy degrees of freedom \cite{APS}-\cite{CKM}.

 A closely related point concerns the phase of the system.   
If the dual group were  in  Higgs phase, the multiplet  structure among the monopoles would get lost, generally.  Therefore one must study the dual  (${\tilde H}$)  
system in  confinement phase.\footnote{
Non-abelian monopoles in the  {\it Coulomb phase}  suffer from the  difficulties already discussed.
} 
 But then, according to the standard electromagnetic duality argument, {\it one  must analyze the  electric  system   in Higgs phase. }  The monopoles will appear confined by the vortices of  the  $H$ system,  which can be naturally interpreted as confining string of the dual system ${\tilde H}$.

We are thus led to study   the system with a hierarchical symmetry breaking,
\be
 G   \,\,\,{\stackrel { v_{1}}  {\longrightarrow}} \,\,\, H  \,\,\,{\stackrel {v_{2}} {\longrightarrow}} \,\,\,
 {\bf 1},  \label{hierarchybis} 
\ee
where
\be   v_{1}  \gg   v_{2},  
\ee
instead of the original system (\ref{this}).
The smaller VEV breaks $H$ completely.   However,  in order for the  degeneracy among the monopoles not to be broken by the breaking   at the scale $v_{2}$, we  require 
that some global color-flavor diagonal group
\be   H_{C+F} \subset   H_{color} \otimes G_{F}   \label{symmetry}
\ee
remains unbroken (see  below).

As we shall see, such a scenario is very naturally realized in ${\cal N}=2$ supersymmetric theories.  
An important lesson one learns from these considerations (and from the explicit models), is 
that the role of the massless flavor is fundamental.      This manifests itself in more than one ways. 
\begin{description}
  \item[(i)] $H$  must be non-asymptotically free,  this requires that there be sufficient number of massless flavors: otherwise, $H$ interactions would  become strong at low energies and $H$ group can break itself dynamically;
  \item[(ii)]  The physics of the $r$ vacua \cite{APS,CKM} indeed shows that the non-Abelian dual group $SU(r)$  appear only for $r \le \frac{N_{f}}{2}$.  This limit can be understood from the renormalization group:  in order for a nontrivial $r$ vacuum to exist, there must be at least  $2\, r $ massless  matter flavor  in the original, electric  theory;
   \item[(iii)]  Non-abelian vortices \cite{HT,ABEKY}, which as we shall see are closely related to the concept of non-Abelian monopoles,
    require also an exact  flavor group. 
The non-Abelian flux moduli arise as a result of an exact  color-flavor diagonal symmetry of the system,  broken by individual soliton vortices.
\end{description}



\section{Non-Abelian monopoles from  vortex moduli  \label{relating}}  

It turns out that the properties of the monopoles induced by the breaking 
\be   G \to H 
\ee
are closely related to  the properties of  the vortices, which develop when the low-energy $H$ gauge theory is put in Higgs phase by a set of scalar VEVs,
$H \to  {\bf 1}$.  The crucial instrument is the 
 exact homotopy sequence, 
\be    \cdots \to   \pi_{2}(G)   \to    \pi_{2}(G/H)   \to      \pi_{1}(H)  \to   \pi_{1} (G) \to \cdots    \label{homotopy}
\ee
But first a few words on homotopy groups and on the use of these relations to characterize the semiclassical monopoles. 
We shall come back to consider  monopole-vortex mixed configurations  later. 

\begin{figure}
\begin{center}
\includegraphics[width=3in]{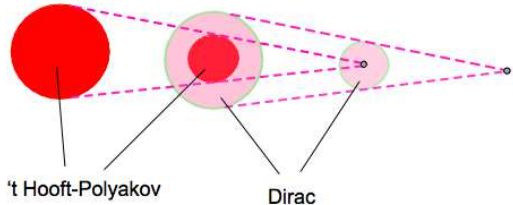}
\caption{\small A pictorial representation of the exact homotopy  
sequence,  (\ref{homotopy}),  with  the  leftmost figure  
corresponding to  $\pi_{2}(G/H)$.} \label{sequence}
\end{center}
\end{figure}

$\pi_{1}(M)$ and   $\pi_{2}(M)$  are the first and second homotopy groups, respectively,   representing the distinct classes of maps from $S^{1}$  or $S^{2}$   to the  (group)  manifold $M$.    Now  ``products'' among such equivalent classes can be defined and  they turn out to form a group  structure  \cite{dubrovin,Coleman}.   The definition of ``the relative homotopy groups'' such as $ \pi_{2}(G/H) $  and  the proof of the exactness of the  sequence (\ref{homotopy}) can be found in the first  reference.   
An exact sequence is a useful tool for studying the structure of different groups through their
correspondences (group homomorphisms).   ``Exact''  means that the kernel of the map at any point of the chain
is equal to the image of  the preceding map.  Such relations are shown pictorially in  Fig. \ref{sequence}.     These sequences can be used, for instance, as follows.  
Assume for simplicity that $\pi_{2}(G) $ and  $\pi_{1} (G) $ are both trivial.   In this case it is clear that each   element of  $\pi_{1} (H) $
is an image of a corresponding element of   $\pi_{2} (G/H)$:   all monopoles are regular, 't Hooft-Polyakov monopoles.

Consider now the case $\pi_{1} (G) $ is nontrivial. Take for concreteness $G=SO(3)$,  with $\pi_{1}(SO(3))={\mathbf Z}_{2}$, and 
$H= U(1)$, with $\pi_{1}(U(1))={\mathbf Z}.$    For any compact Lie groups $\pi_{2}(G) ={\bf 1}$.    The exact sequence illustrated in  Fig. \ref{sequence} in this case implies that the monopoles, classified by $\pi_{1}(U(1))={\mathbf Z}$  can further by divided into 
two classes, one belonging to the image of $\pi_{2}(SO(3)/U(1))$ -- 't Hooft-Polyakov monopoles! -- and  those which are not related to the breaking -- the singular, Dirac monopoles.  The correspondence is two-to-one: the monopoles of magnetic charges  $2\, n$ times ($n=1,2,\ldots$)  the Dirac unit are regular monopoles while those with charges $2\, n +1$ are Dirac monopoles. 
In other words,  the regular monopoles correspond to the  kernel of the map $  \pi_{1}(H)  \to   \pi_{1} (G) $  (Coleman \cite{Coleman}). 

\begin{figure}
\begin{center}
\includegraphics[width=2in]{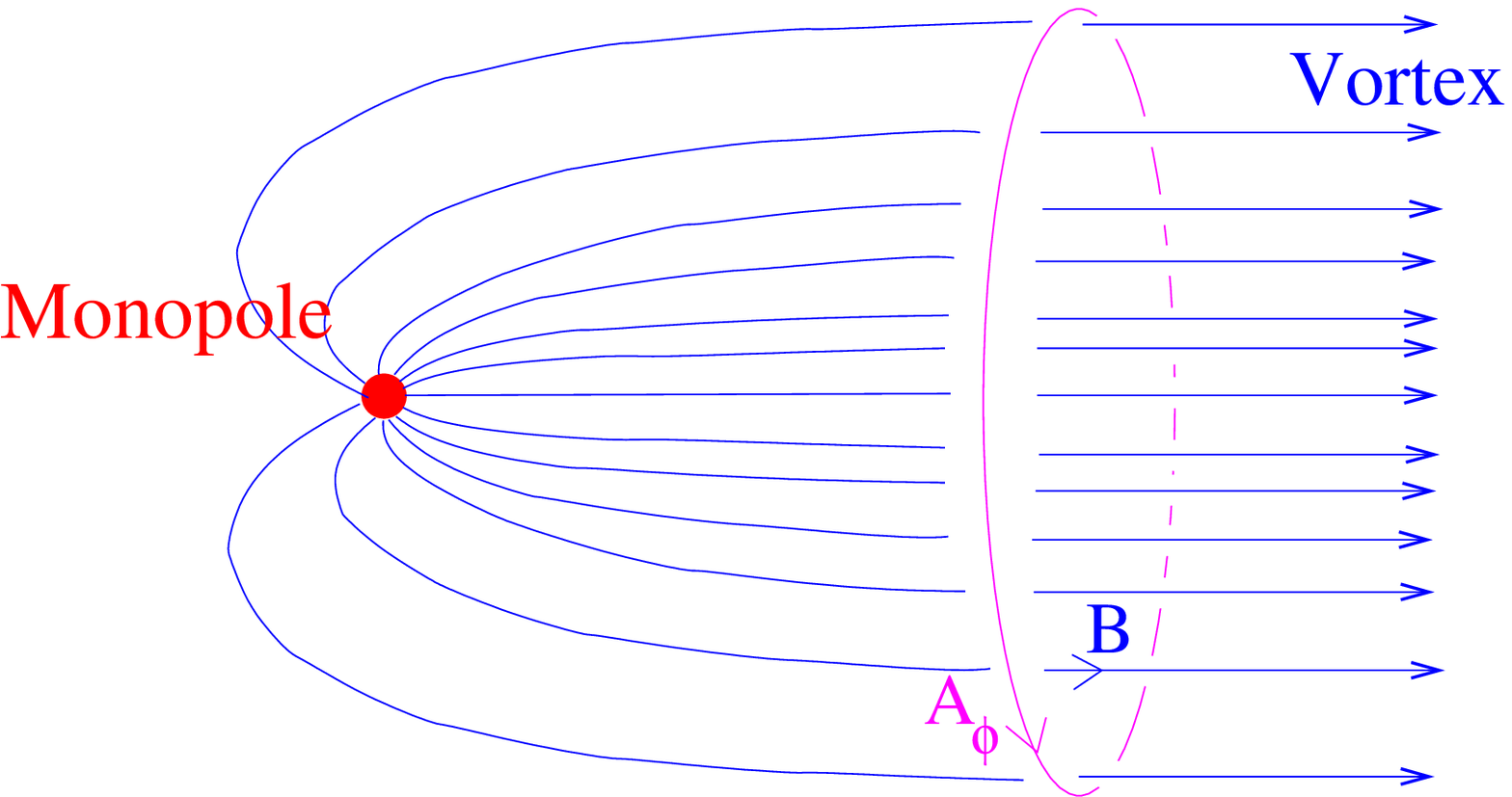}
\caption{ }
\label{monovortex}
\end{center}
\end{figure}

The exact sequence (\ref{homotopy}) assumes  an important significance when we consider the system with
a hierarchical symmetry breaking (\ref{hierarchybis}),
\[
 G   \,\,\,{\stackrel { v_{1}}  {\longrightarrow}} \,\,\, H  \,\,\,{\stackrel {v_{2}} {\longrightarrow}} \,\,\,
 {\bf 1}.
\]
   As $H$ is now completely broken the low-energy theory has vortices, classified by $\pi_{1}(H)$.  
If $\pi_{1}(G) ={\bf 1}$, however,  the full theory cannot have vortices.  This apparent paradox is solved when one realizes that there is another  related paradox:   monopoles representing $\pi_{2}(G/H)$ cannot be stable,   because in the full  theory  the gauge group is completely broken, $G \to 
{\bf 1}$,  and because  for any Lie group, $\pi_{2}(G)={\bf 1}.$   These paradoxes solve themselves:   the  vortices of the low-energy theory end at the monopoles, which have large but finite masses.  Or they are broken in the middle by   (though suppressed)
monopole-antimonopole pair production.   Vice versa,  the monopoles are not stable  as its  flux is carried  away by the vortex.  See Fig. \ref{monovortex}

Applied to the case  of $SO(3) \to U(1) \to {\bf 1}$, 
this was precisely the logic used  by 't Hooft in his pioneering paper on the monopoles.  As is seen from Fig. \ref{sequence},  the  vortices 
($\pi_{1}(U(1))=\mathbf Z$)  of the winding number two,    corresponding to the trivial element of   $\pi_{1}(SO(3)) =  {\mathbf Z}_{2}$,     should not be stable  in the full theory:  there must be a regular monopole-like configuration, having the magnetic charge twice the Dirac unit, $g_{m} = {4 \pi}/{g}$, where $g$  is the  the gauge coupling constant of the $SO(3)$ theory, acting as a source or a sink of the magnetic flux  (Fig. \ref{monovortex}). \footnote{The relation appears  to violate the Dirac quantization condition: actually
the minimum electric charge which could be introduced  in the theory is that of a quark, $e= g/2$,  and which satisfies
$ g_{m} \, e = 2 \pi$, in accordance  with  Dirac's condition. }     

An important new aspect we have here,  as compared to the case discussed by 't Hooft \cite{TH}  is 
 that now the unbroken group $H$ is non-Abelian and
that the low-energy vortices carry continuous, non-Abelian flux moduli.  
As the color-flavor diagonal  symmetry $H_{C+F}$   is an exact unbroken symmetry of the full theory, and the non-Abelian moduli among the low-energy vortices is a consequence of it,  it follows that the   the monopoles appearing as the endpoints of such vortices
 carry the same continuous moduli.

 \begin{center}
{\tt    The monopole transformation properties follow from those of the 
 vortices, which can be studied exactly  in the low-energy approximation.    }
 \end{center}

\section {${\cal N}=2$ supersymmetric gauge theories and light non-Abelian monopoles}   

It is always a  healthy attitude to try to test one's general  idea  against a concrete model.   For various reasons 
it turns out that ${\cal N}=2$ models 
provides  a good  testing ground, as  the results of strong infrared dynamics are known in the  form of exact Seiberg-Witten curves.
Another advantage is that by varying certain parameters upon which the system depends holomorphically,  as  is
usual in supersymmetric theories,  one can study the system  Eq. (\ref{hierarchy}) in different regimes.  

In the regions of parameters where  $v_{1}\gg v_{2}\gg \Lambda$, 
semiclassical analysis in the original  electric theory  is justified, and one can study monopoles  (in the effective theory at mass scales much higher than $v_{2}$)
and  separately, the vortices (in the effective theory valid  at mass scales much lower  than $v_{1}$).  The symmetry  and  homotopy-map  argument allows to obtain the   missing information about the non-Abelian transformation properties of the monopoles, from the known properties of the vortices.    
We come back to this discussion in Section \ref{sec:duality}.  In the concrete models studied there the breaking mass scales are given by
$ m_{i}=m   \sim  v_{1}$;   $\sqrt{\mu \, m} \sim  v_{2}$,  so  the parameter regions explored correspond to 
$|m_{i}| \gg |\mu|  \gg \Lambda$.   

These results are then checked against the fully quantum mechanical  results on the monopoles appearing as the massless degrees of freedom 
in the magnetic dual theory, in the region 
 $v_{1}\sim v_{2} \sim \Lambda$.  This regime will be discussed first.
  In the following   Section \ref{withquarks}, in fact, the parameters are chosen to be   $m_{i}, \mu \sim \Lambda $, and in particular,  $m_{i} \to m$.

  We shall return later (Section \ref{sec:duality}) to see that  how our  ideas on non-Abelian duality  based on the   hierarchical symmetry breaking and on color-flavor diagonal symmetry  can be studied in the same  model 
  reliably    and see that the results found match the full quantum results.   
 
\subsection{Seiberg-Witten solution of  pure ${\cal N}=2$  Yang-Mills} 


${\cal N}=2$ supersymmetric   $SU(2)$  Yang-Mills theory is  described by the Lagrangian,  
\be
L=     \frac{1}{ 8 \pi} \Im \, \tau_{cl} \left[\int d^4 \theta \,
\Phi^{\dagger} e^V \Phi +\int d^2 \tht\,\frac{1}{ 2} W W\right]
\label{susylagrangian}
\ee
where
\be
\tau_{cl} \equiv  \frac{\theta_0 }{ 2  \pi} + \frac{4 \pi i }{ g_0^2}
\ee
is the bare $\theta$ parameter and coupling constant.
 $\Phi= \phi \, + \, \sqrt2 \,
\theta \,\psi + \, \ldots \, $, and $W_{\alpha} = -i \lambda \, + \, \frac{i
}{ 2} \, (\sigma^{\mu} \, {\bar \sigma}^{\nu})_{\alpha}^{\beta} \, F_{\mu \nu} \,
\theta_{\beta} + \, \ldots $ are   ${\cal N}=1$ chiral and gauge superfields,  both in the adjoint representation of
the gauge group.   The theory possesses  ${\cal N}=2$ supersymmetry as there are two gauginos,  $\lambda$ and $\psi$.  

The scalar potential in this case is just the  so-called $D$ term 
\be    V_{D} =   \frac{g^{2}}{8} |  [\Phi^{\dagger},  \Phi] |^{2},   
\ee
only, and the system  has a continuous vacuum degeneracy  (CMS- classical moduli  space), parametrized by a complex number $a$, 
\be    \brc \Phi \ckt =  \left(\begin{array}{cc}a & 0 \\0 & -a\end{array}\right). 
\ee
At any given $a$ the gauge symmetry is broken by Higgs mechanism to $U(1)$.  The low energy theory is a $U(1)$ theory, describing the photon and photino $\lambda$, and the ${\cal N}=2$ partners,   $A= (A, \psi)$.

The general requirement of  ${\cal N}=2$  supersymmetry implies  that the Lagrangian has the form, 
\be    L_{eff}=   \frac{1}{4\pi}   \Im \, [ \int d^{4} \theta \frac{d F(A)}{d  A}    \, {\bar A}  + \int  \frac{1}{2}  \frac {d^{2}F(A)}{d A^{2}}   \, W^{\alpha}\, W_{\alpha}\, ],    \label{U1}
\ee
with $F(A)$  is  holomorphic in $A$.  $F(A)$  is known as prepotential.    Going to component fields, the fermionic and gauge parts take the form, 
\[  L_{ferm} =  \frac{1}{8 \pi^{2}}  [ \Im \frac {d^{2}F(A)}{d A^{2}}]  ( i {\bar \psi}  {\bar \sigma}^{\mu}  {\bar D}_{\mu} \psi  +   i {\bar \lambda}  {\bar \sigma}^{\mu}  {\bar D}_{\mu} \lambda + \ldots ),  
\]
\[  L_{gauge} =  \frac{1}{16 \pi^{2}}  [ \Im \frac {d^{2}F(A)}{d A^{2}}]  (   F_{\mu \nu}^{2} +  i  \, F_{\mu \nu}\,{\tilde F}^{\mu \nu} + \ldots   ),
\]
which shows clearly  $\psi$ and $\lambda$  have the same properties as the adjoint fermions  ($SU_{R}(2)$ global symmetry of ${\cal N}=2$  supersymmetry);  the second formula shows that 
\[  \tau_{eff} =   \frac{d A_{D}}{dA} =   \frac {d^{2}F(A)}{d A^{2}},  \qquad    A_{D} \equiv   \frac{d F(A)}{d  A}, 
\]
acts as the low-energy effective (complex) coupling constant
\be    \tau_{eff} =     \frac{\theta_{eff}}{2\pi}  + \frac{4\pi i}{g_{eff}^{2}},  
\label{taueff}\ee

Let us recall that in general  $4D$   supersymmetric sigma model, with a set of scalar multilpets $\Phi$, the kinetic term is given by a  (real)  K\"ahler potential  
\[  L =  \int d^{4}\theta   \, K(\Phi,  {\bar \Phi}) =   \frac{\de^{2}K}{\de \phi_{i} \, \de {\bar \phi}_{j}} \,
\de_{\mu} \phi_{i}\, \de^{\mu} {\bar \phi}_{j} + \ldots.  
\]
Here  the K\"ahler potential  has a special form, determined by  the prepotential,  
\[   K=  \frac{1}{2 i}  [     \, \frac{d F(A)}{d  A_{i}}  \, {\bar A}_{i}  -   \frac{d F({\bar A} )}{d  {\bar A}_{i}}  \, {A}_{i}\, ]
\]
(termed special  geometry). 

Coming back to the $SU(2)$  ${\cal N}=2$ Yang-Mills theory where there is only one scalar multiplet $A$,    
the bosonic part of the Lagrangian has the form, 
\[   L_{bos} =   \frac{1}{2 \, i} \, (\de_{\mu} \, a_{D}  \, \de^{\mu}  {\bar a}  - \de_{\mu} \, a  \, \de^{\mu}  {\bar a}_{D}  )  
+  \Im \tau(a)  ( F_{\mu \nu}^{+} )^{2},  
\qquad   F_{\mu \nu}^{+}=  F_{\mu \nu}+   i\, {\tilde F}_{\mu \nu}. 
\]
 Now this model has a nice property of (form) invariance  under the generalized electromagnetic duality transformation \cite{GZ} 
\be     \left(\begin{array}{c}a_D \\a\end{array}\right)  \to   M\, \left(\begin{array}{c}a_D \\a\end{array}\right), 
\qquad      \left(\begin{array}{c}F_{\mu \nu}^+  \\  G_{\mu \nu}^+   \end{array}\right)   \to   M\, \left(\begin{array}{c} F_{\mu \nu}^+ \\ G_{\mu \nu}^+\end{array}\right); 
 \label{SL2Z}\ee
where 
\[   G_{\mu \nu}^+   \equiv \frac{1}{2} \frac{\de }{\de F_{\mu \nu}^{+}}    \, [ \tau(a) \, F_{\mu \nu}^{+\, {2} } ] 
\]
and $M$  is an  $SL(2, Z)$   matrix,    
\[  M=    \left(\begin{array}{cc}A & B \\C & D\end{array}\right), \qquad   A\,D -  B\, C
=1.   \]

Such an invariance group  includes the electromagnetic duality transformation $F_{\mu \nu} \leftrightarrow  {\tilde F}_{\mu \nu}$,
together with  $a  \leftrightarrow a_{D}$.   

Since $F(A)$  is holomorphic,  so is $\tau(A)$:  it is harmonic, $\nabla \tau = \nabla  \Im \tau =0. $    Thus $\Im \tau$ cannot be everywhere positive.  This means that $A$ cannot be a good global  variable everywhere in the field space: there must be some singularities where 
the description in terms of $a$,  $ F_{\mu \nu} $ fails. 
    
\begin{figure}
\begin{center}
\includegraphics[width=2.5 in]{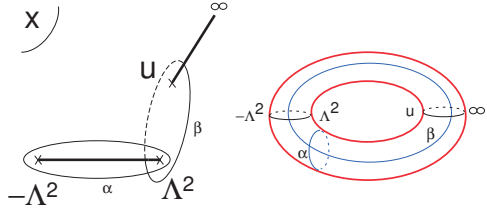}
\caption{\small  The torus (\ref{torus}) represented as a two-sheeted Riemann surfaces,  with two branch cuts (left).  Note that two Riemann spheres attached at two cuts are equivalent to a  torus (figure on the right).    }
\label{cycles}
\end{center}
\end{figure}

The beautiful  argument  by Seiberg and Witten \cite{SW1,SW2} that the singularity  be related to  the point where  the magnetic monopole of the theory  --  as the bosonic part of the model is just the Giorgi-Glashow model  the soliton monopoles found  by  't  Hooft and Polyakov are part of the spectrum  --   becomes {\it massless}  due to quantum effects, and the consequent determination of the  the prepotential $F(A)$  are by now well known.   For completeness we summarize the main points of the solution in Appendix \ref{SWcurve}. 
Let us recall  the main result here:  by introducing an auxiliary  torus  (whose genus $1$  corresponds to the rank of the gauge group $SU(2)$), described by the algebraic {\it curve}
\be   y^{2} =   (x^{2}-\Lambda^{4}) (x-u) =  (x+\Lambda^{2})   (x-\Lambda^{2}) (x- u),  \qquad  u\equiv \brc \Tr \Phi^{2} \ckt,  
\label{torus}\ee  
the solution is expressed as 
\be    \frac{d a_{D}}{d u}  =  \oint_{\beta}    \frac{dx}{y}, \qquad    \frac{d a} {d u}  =  \oint_{\alpha}    \frac{dx}{y}, 
\label{keystep}
\ee
where $\alpha$ and $\beta$ are the two canonical cycles on the torus,  Fig. \ref{cycles}.  
Explicitly, 
\[   a_{D}(u) =  \frac{\sqrt 2}{\pi}  \int_{\Lambda^{2}}^{u}   \left(\frac{x-u}{x^{2}- \Lambda^{4}}\right)^{1/2}= i\, \frac{u- \Lambda^{2}}{2} \,F(\frac{1}{2}, \frac{1}{2}; 2;   \frac{\Lambda^{2}-u}{2} ), 
\]
\be   a(u) =  \frac{\sqrt 2}{\pi}  \int_{\Lambda^{2}}^{\Lambda^{2}}   \left(\frac{x-u}{x^{2}- \Lambda^{4}}\right)^{1/2}= \sqrt{2}  \, (u+ \Lambda^{2})^{1/2}\, F(- \frac{1}{2}, \frac{1}{2}; 1;   \frac{2}{u + \Lambda^{2}} ). \label{SWsol}
\ee
The key step of the solution Eq. (\ref{keystep}) was the theorem in algebraic geometry  that the integrals of the  
holomorphic  differential   ($\frac{dx}{y}$  in our case of the genus one torus (\ref{torus}))   
along the canonical cycles   $\alpha$ and $\beta$   (they are called period integrals) satisfy
\[    \Im    \frac{    \oint_{\alpha}    \frac{dx}{y}}  { \oint_{\beta}   \frac{dx}{y}} >0, 
\] 
independently  of the way canonical cycles are redefined.   According to   the identification of the period integrals with the physical  
quantities as Eq. (\ref{keystep}) this guarantees that 
\[    \Im  \,   \tau_{eff}  =  \Im    \frac{d a_{D}}{da }  =   \frac{4\pi}{g_{eff}^{2} }   >0.        \]
Let us add  several remarks.
\begin{description}
\item[(i)]     Another key observation by Seiberg-Witten is that the ${\cal N}=2$  supersymmetry  implies  an exact mass formula 
for   BPS saturated states with magnetic and electric charges $n_{m}, n_{e}$:
\be    M_{n_{m}, n_{e}}  = \sqrt{2}  \, |  n_{m}\, a_{D} +   n_{e}\, a|.  \label{exactmass}
\ee
  This is a  consequence of the fact that the system has an underlying ${\cal N}=2$  supersymmetry with a central extension. See Appendix \ref{susyalgebra}. 
   This formula generalizes  the standard Higgs formula,  $M_{0, n_{e}}= g\, n_{e} \brc  \phi  \ckt$,   as $ a  \sim  g \brc  \phi  \ckt$ semiclassically, 
   and  at the same time,   the 't Hooft-Polyakov monopole mass formula, 
   $M_{n_{m}, 0} =  4\,\pi \, n_{m}\, \brc  \phi  \ckt /  g  $    (semiclassically  $a_{D}\sim     4\,\pi   \, \brc  \phi  \ckt /  g $).    Note that in the fully quantum formula (\ref{exactmass}) the magnetic and electric charges appear  symmetrically.      Indeed  the mass formula is invariant under  the generalized duality transformations  (\ref{SL2Z}),  modulo appropriate relabeling of magnetic and electric charges. 
      
  \item[(ii)]   Quite remarkably the low-energy effective action thus determined contains quantum effects in its entirety, the one-loop perturbative effects plus the  sum of infinite instanton contributions.
   Indeed, the Seiberg-Witten curves have been checked against direct instanton calculations \cite{Instcheck}, and more recently, have been 
rederived by an explicit instanton resummation \cite{Nekr}.

\item[(iv)]  The Seiberg-Witten solution  nicely solves an old (apparent) paradox related to the Dirac quantization versus renormalization group \cite{Coleman}:  how can  the relation  $ g_{m} \, g_{e} =  2 \,\pi\, n$,  $n=0, 1, \ldots$  be  compatible with the fact that both the electric and magnetic  charges are Abelian $U(1)$ coupling constants,  expected to get renormalized in the same direction?  In the Seiberg-Witten solution,  $g_{m}(\mu)$ gets renormalized as in (magnetic version of) QED, through monopole loops, with monopoles replacing  the role of the electron.  The same infrared behavior is explained, in the original electric picture,  as due to  instanton-induced nonperturbative renormalization of the electric coupling constant  $g_{e}(\mu)$. As a consequence $ g_{m}(\mu) \, g_{e}(\mu)=  4\, \pi$  holds \cite{SW1}   at any infrared cutoff $\mu =  \brc a_{D} \ckt.$ 
For other subtle issues related to renormalization group properties of Seiberg-Witten solution, see \cite{KK}.

\item[(iv)]  How do we know that these  massless monopoles  are  related to the 't Hooft-Polyakov monopoles?    That they {\it are} 
indeed them,  can be verified by studying the  electric and quark (in the cases with $N_{f}=1,2,3$) number charges. As is well known the 't Hooft-Polyakov monopoles 
acquire these $U(1)$  charges quantum mechanically, via  a  beautiful phenomenon of charge fractionalization \cite{GoldWilc},  which  in this specific situation are the  Witten's \cite{Witten} and  Jackiw-Rebbi's   effects \cite{JR}.  By moving within the space of vacua (QMS) and going into the regions where semiclassical approximation is valid 
(where  $u= \brc \Tr \, \Phi^{2} \ckt \gg \Lambda^{2}$), one  can compare these fractional  $U(1)$ charges read off from the leading terms of the exact Seiberg-Witten solution  with the ones obtained many years earlier by  standard quantization of fermion fields  around the semiclassical monopole backgrounds \cite{NiemSeme}. The results exactly match \cite{FF,KT,Rebhan}. 

  \item[(iv)]   The low-energy effective Lagrangian near one of the singularities, {\it e.g.},   $u = \Lambda^{2}$,   
  looks like a (dual) QED  with a massless monopole,  whose Lagrangian has the standard ${\cal N}=2$  QED form, 
\bqa  L &=&    \frac{1}{4\pi}   \Im \, [ \int d^{4} \theta \frac{d F(A_{D})}{d  A_{D}}     \, {\bar A_{D}}  + \int  \frac{1}{2}  \frac {d^{2}F(A_{D})}{d A_{D}^{2}}   \, W_{D}^{\alpha} \, W_{D\, \alpha}\, ]   + \non \\
&&   + \int d^4 \theta  ({\bar M }  e^{V_{D}}  M  +   {\tilde M}   e^{- V_{D}}  {\bar {\tilde M}} )
+  \int d^{2}\theta \sqrt{2} {\tilde M} A_{D} M, 
 \label{seethis} \eea 
  where the gauge terms are just the dual of Eq. (\ref{U1});  the third and fourth terms describe the monopole.   
   \item[(v)]     Addition of a ${\cal N}=1$  perturbation, the adjoint scalar mass term, $\mu \, \Tr \Phi^{2}$  in the original electric theory    induces $\Delta L=   \mu \, U(A_{D}), $  where the function $U(A_{D})$ is the inverse of the 
  solution $a_{D}(u)$.  By minimizing the potential,    the degeneracy (quantum moduli space -- QMS) is eliminated leaving just two vacua, where   
  \[   a_{D}=0, \quad  u= \brc \Tr \Phi^{2}  \ckt = \pm \Lambda^{2}, \qquad     \brc M \ckt =  \brc {\tilde M} \ckt = \mu \, \frac{\de U }{\de A_{D}} \sim \mu \, \Lambda.  
  \]
  The first  result says that the magnetic monopole is massless in this vacuum (see Eq. (\ref{seethis})), the third states that 
 the magnetic monopole condenses,  leading to confinement \`a la 't Hooft-Mandelstam.  This is perhaps the first example of nontrivial 4D system where this phenomenon  has been demonstrated explicitly and  analytically. 
\end{description}

\subsection{Seiberg-Witten solutions for ${\cal N}=2$  models with quarks   \label{withquarks}} 

   A general enthusiasm (alarm?) caused by the news that the $SU(2)$  Seiberg-Witten model   with  a small ${\cal N}=1$
perturbation  exhibited  the  't Hooft-Mandelstam mechanism of confinement,    
was followed by a widespread delusion (relief?) among theoretical physicists  when it was realized that the light  monopoles appearing in the low-energy  theory  were  Abelian and at the same time  confinement was accompanied by dynamical Abelianization.   This surely was not a good  model of QCD!   The fact that in the $SU(2)$ models with  $N_{f}=1,2,3$  hypermultiplets of quarks, studied in  the  (quite remarkable)  second paper by Seiberg and Witten \cite{SW2}, 
as well as in pure ${\cal N}=2$  Yang-Mills theories with more general gauge groups \cite{curves},  
the low-energy  monopoles  were always Abelian, did not help.

  What was not realized at the time, however,    was the fact that there was a clear reason for the Abelianization in these simplest models (see Subsection
  \ref{sec:exactqb}    below), 
and  that, in the context of a  more general  class of ${\cal N}=2$ theories with quark multiplets, 
Abelian  confinement  belonged to the exceptional cases. In fact,  
confinement is more typically  caused by  condensation of non-Abelian monopoles, as the subsequent analyses have revealed. 
We shall below briefly summarize the main features of these models, with technical aspects kept  at its minimum.

  
  The systems we consider are simple generalization of the ${\cal N}=2$  models with ``quark'' multiplets.
  The $N=1$ chiral and gauge superfields $\Phi= \phi \, + \, \sqrt2 \,
\theta\,\psi + \, \ldots \, $, and $W_{\alpha} = -i \lambda \, + \, \frac{i
}{ 2} \, (\sigma^{\mu} \, {\bar \sigma}^{\nu})_{\alpha}^{\beta} \, F_{\mu \nu} \,
\theta_{\beta} + \, \ldots $ are both in the adjoint representation of
the gauge group, while the hypermultiplets are taken in the
fundamental representation of the gauge group.  The Lagrangian takes the form, 
 \be
L=     \frac{1}{ 8 \pi} \Im \, \tau_{cl} \left[\int d^4 \theta\,
\Phi^{\dagger} e^V \Phi +\int d^2 \theta\,\frac{1}{ 2} W W\right]
+ L^{(quarks)} + \Delta L +  \Delta^{\prime} L,   \label{lagrangianGeneral}
\ee
\be L^{(quarks)}= \sum_i \, [ \int d^4 \theta\, \{ Q_i^{\dagger} e^V
Q_i + {\tilde Q_i}^{\dagger}  e^{ {\tilde V}}    {\tilde Q}_i \} +
\int d^2 \theta
\, \{ \sqrt{2} {\tilde Q}_i \Phi Q^i    +      m_i   {\tilde Q}_i    Q^i   \}
\label{lagquark}
\ee
describes the $n_{f}$ flavors of hypermultiplets (``quarks''),  
\be
\tau_{cl} \equiv  \frac{\theta_0 }{ \pi} + \frac{8 \pi i }{ g_0^2}
\label{struc}
\ee
is the bare $\theta$ parameter and coupling constant.
The $N=1$ chiral and gauge superfields $\Phi= \phi \, + \, \sqrt2 \,
\theta\,\psi + \, \ldots \, $, and $W_{\alpha} = -i \lambda \, + \, \frac{i
}{ 2} \, (\sigma^{\mu} \, {\bar \sigma}^{\nu})_{\alpha}^{\beta} \, F_{\mu \nu} \,\theta_{\beta} + \, \ldots $ are both in the adjoint representation of
the gauge group, while the hypermultiplets are taken in the
fundamental representation of the gauge group.

We    consider      small {\it
generic}      nonvanishing
   bare masses   $m_{i}$   for the hypermultiplets
(``quarks''), which is consistent with ${\cal N}=2$ supersymmetry.    
Furthermore   it is convenient to introduce the mass for the adjoint scalar multiplet 
\be
\Delta   L=   \int \, d^2 \theta \,\mu  \,\Tr \, \Phi^2
\label{N1pert}
\ee
which breaks supersymmetry to ${\cal N}=1$.   
    An advantage of doing so  is that
    all flat directions   are eliminated  and one is left
with a finite number of isolated vacua;   keeping track
of this number  (and the symmetry breaking pattern in each of them)  allows us to make    highly nontrivial check of our
analyses at various stages.

Below we  summarize the physical results on these systems.    To solve the system (\ref{lagrangianGeneral})  the first step is the generalization of the curve Eq. (\ref{torus})  to the case of general group $G$.  When the  breaking is maximum, $G \to  U(1)^{r_{G}}$  where $r_{G}$ is the rank of the group $G$, we set $\mu =0$ and 
consider vacua   
\be   \brc \Phi \ckt = \diag  ( \phi_{1}, \phi_{2}, \ldots ), \qquad    \phi_{1} \ne  \phi_{2}, \quad etc.
\ee
The auxiliary genus $g= N_c-1$ (or   $N_{c}$) curves for
$SU(N_c)$ ($USp(2 N_c)$) theories corresponding to these classical vacua  (called Coulomb branch of the moduli space)    are given by
\begin{equation}
    y^{2} = \prod_{k=1}^{n_{c}}(x-\phi_{k})^{2} + 4 \Lambda^{2n_{c}-n_{f}}
    \prod_{j=1}^{n_{f}}(x+m_{j}), \qquad   SU(N_c), \, \,\,  N_f \le 2N_c-2,
\label{curve1} \end{equation}
and
\begin{equation}
    y^{2} = \prod_{k=1}^{n_{c}}(x-\phi_{k})^{2} + 4 \Lambda
    \prod_{j=1}^{n_{f}}\left(x+m_{j} + \frac{\Lambda }{ N_c} \right), \qquad
SU(N_c),
\, \,\,  N_f= 2N_c-1,
\label{curve2} \end{equation}
with $\phi_{k}$ subject to the constraint $\sum_{k=1}^{n_{c}}\phi_{k} =
0$, and
\begin{equation}
    x y^{2} = \left[ x \prod_{a=1}^{n_{c}} (x-\phi_{a}^{2})^{2}
      + 2 \Lambda^{2n_{c}+2-n_{f}} m_{1} \cdots m_{n_{f}} \right]^{2}
    - 4 \Lambda^{2(2n_{c}+2-n_{f})} \prod_{i=1}^{n_{f}}(x+m_{i}^{2})  \label{curve3}
\end{equation}
for  $USp(2 N_c)$.   Analogous results for $SO(N_{c})$ theories are also known. 

The connection between these genus $g$ hypertori and physics  is
made \cite{SW1}-\cite{HO}   through the
identification of various   period integrals of   the { holomorphic
differentials}
on the curves with   $(d a_{D i} / du_j, d a_{ i} / du_j)$,   where
the gauge invariant parameters
$u_j$'s are defined by the standard relation,
\be   \prod_{a=1}^{n_{c}}(x-\phi_{a})  =  \sum_{k=0}^{N_c}  u_k  \,
x^{N_c-k}, \qquad  u_0=1, \quad u_1=0,  \qquad
SU(N_c);
\ee 
\be   \prod_{a=1}^{n_{c}}(x-\phi_{a}^2)  =  \sum_{k=0}^{N_c}  u_k  \,
x^{N_c-k}, \qquad  u_0=1, \qquad  USp(2 N_c),   \ee 
and  $u_2\equiv \brc \Tr \, \Phi^2 \ckt, $    $u_3 \equiv \brc \Tr \, \Phi^3 \ckt, $ etc.     The
VEVS of $a_{Di}, \,\, a_{i}$, which are    directly related to the
physical  masses of the
BPS  particles through the exact  Seiberg-Witten mass formula \cite{SW1,SW2}
\be   M^{n_{mi}, n_{ei}, S_k} = \sqrt2   \,  \left|  \sum_{i=1}^g  (
n_{mi} \, a_{Di} + n_{ei}  \,  a_{i}  )     +   \sum_k   S_k   m_k
\right|, \ee 
   are  constructed as   integrals   over the non-trivial cycles   of the meromorphic
differentials  on the  curves.  $S_{k}$ are the $i$-th quark number  charge of the monopole under consideration, 
which enters the formula for the central charges (hence the mass). 

\begin{description}
  \item[(i)]  These formulae naturally generalize those of the pure $SU(2)$ theory, Eq. (\ref{keystep}), Eq. (\ref{exactmass}).   The singularities of the curves Eq. (\ref{curve1})-Eq. (\ref{curve3}) are the points in the space of vacua  (QMS)  where various  particles become massless. 
  \item[(ii)] When  $m_{i} \gg \Lambda$ these singularities are at the points  where 
 $\phi\sim m_{i}$  (where the quarks become massless -- see Eq. (\ref{lagquark})) and at the points where monopoles of pure Yang-Mills theory become massless.  The latter are the 
 points  the curve of the Yang-Mills theory, 
 \[  y^{2} = \prod_{k=1}^{n_{c}}(x-\phi_{k})^{2} + 4 \Lambda_{YM}^{2n_{c}}
 \]
  become maximally singular,   $\sim  \prod_{i=1}^{n_{c}-1}  (x- x_{i})^{2}  (x-\alpha) (x-\beta)$.  
 \item[(iii)] It is the property of these curves that when $m_{i} \sim   \Lambda$  all singularities  are found to correspond to magnetic degrees of freedom 
 (massless monopoles and dyons).    To trace how, as $m_{i}$ are varied,  the original ``electric'' singularities (massless quarks) make a metamorphosis into magnetic monopoles,  
 due to the movement of certain branch points (or branch cuts) sliding  under other branch cuts (branch surfaces), is a rather complicated business, and has been analyzed satisfactorily  only in the $SU(2)$ theories with matter \cite{SW2,CV}.    
   \item[(iv)] The particular form of the curve specific to different groups reflect different global symmetries.  A nice discussion is given in \cite{curvesbis}. 
\end{description}

\subsection{Exact quantum behavior of light non-Abelian mono\-poles \label{sec:exactqb}}

Physics of confining vacua and properties of light monopoles  in these theories  are studied by identifying all of the ${\cal N}=1$  vacua
(the points in the QMS -- quantum moduli space, that is,  the space of vacua --  which survive the 
${\cal N}=1$ perturbation)  and studying the low-energy action for each of them. The underlying ${\cal N}=2 $ 
 theory, especially with  $m_{i}=0$  or with equal 
masses $m_{i}=m$, has a large continuous degeneracy of vacua (flat directions),  which has been studied by using 
the Seiberg-Witten curves,  non-renormalization of Higgs branch metrics, superconformal points and their universality,
 their moduli structure and 
symmetries,  etc \cite{APS,HO}.   For the purpose of this section, however, we are most interested in the set of vacua which are picked
up when the small generic bare quark masses $m_{i}$ and a small nonzero adjoint mass $\mu$ are present.   
\begin{figure}
\begin{center}
\includegraphics[width=2.5 in]{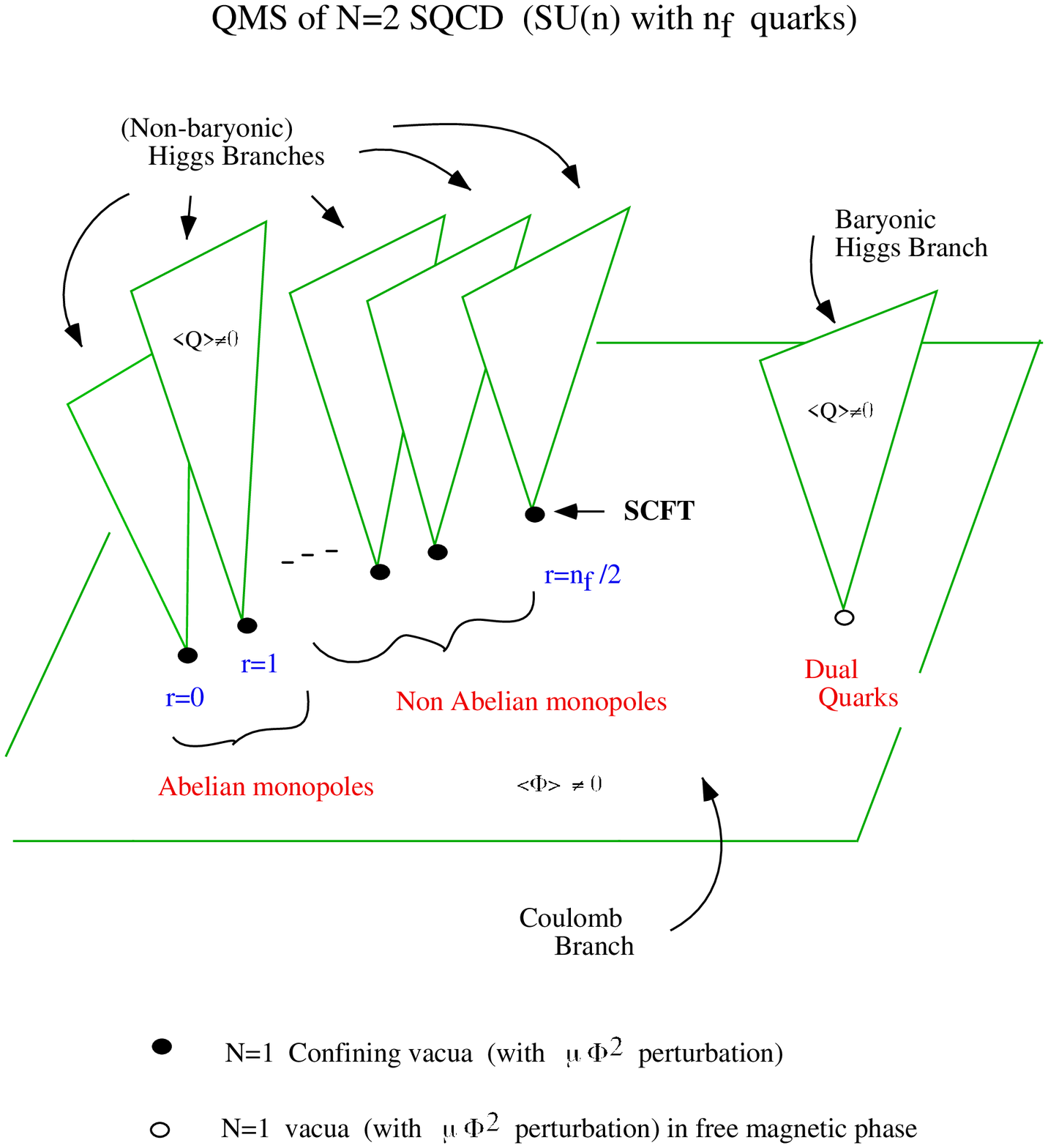}
\caption{ }
\label{rvacua}
\end{center}
\end{figure}
At the roots of these different branches of ${\cal N}=2 $ vacua   where the Higgs branches meet the Coulomb branch, 
 lie  all these   vacua,   which survives the ${\cal N}=1$ perturbation,  Eq. (\ref{N1pert}).  
In $SU(N_{c})$ theories with $N_{f}$ flavors with generic masses,  all ${\cal N}=1$  vacua  arising this way  have been completely classified
\cite{CKM,CKKM}.

For nearly equal   quark masses they fall into classes $r=0,1,\ldots, \frac{N_{f}}{2}$  groups of vacua  near  the ``root of non-baryonic Higgs branches'', and 
for $N_{f} \ge N_{c}$,  there are special vacua at the  ``roots of baryonic Higgs branches''.  These names reflect the fact that in the respective Higgs branch
non-baryonic or  baryonic  squark VEVS, 
\be    \brc   Q_{i}^{a} \, {\tilde Q}_{a}^{j} \ckt,  \qquad  \brc  \epsilon_{a_{1}\,a_{2}\, \ldots,  a_{N_{c}}} Q_{i_{1}}^{a_{1}} Q_{i_{2}}^{a_{2}} \ldots 
Q_{i_{N_{c}}}^{a_{N_{c}}}\ckt,
\ee 
are formed.   See Fig. \ref{rvacua}.   Each group of vacua coalesce in single vacua where the gauge symmetry is enhanced into non-Abelian gauge groups,  as in Table \ref{tabnonb}, taken from Argyres  et. al. \cite{APS}.

   The vacua  at the root of the baryonic branch  are in ``free-magnetic'' phase;  the light non-Abelian magnetic monopoles appear as 
   asymptotic states;  they do not condense,  no confinement and no symmetry-breaking occur.  Although the appearance of the Seiberg dual gauge group, $SU({\tilde N}_{c})$, ${\tilde N}_{c} \equiv   N_{f}- N_{c}$ is certainly intriguing \cite{APS},  these are not type of vacua we are interested in.

   Our main interest is the first classes of  the so-called   ``$r$-vacua'',  where the magnetic gauge group is   
   \[  U(r) \times  U(1)^{N_{c}-r},  
   \]
   and the massless  matter multiplets consist of  $N_{f}$  monopoles in the fundamental representation of $U(r)$, and 
   flavor-singlet Abelian monopoles carrying a single charge, each with respect to one of the $U(1)$ factors   (Table \ref{tabnonb})
   \footnote{We shall use the notation $N_{c}=n_{c}$ indistinguishably, and analogously  $N_{f}=n_{f}$.}.
  
   \begin{table}[b]
\begin{center}
\vskip .3cm
\begin{tabular}{ccccccc}

&   $SU(r)  $     &     $U(1)_0$    &      $ U(1)_1$
&     $\ldots $      &   $U(1)_{n_c-r-1}$    &  $ U(1)_B  $  \\
\hline
$n_f \times  q$     &    ${\underline {\bf r}} $    &     $1$
&     $0$
&      $\ldots$      &     $0$             &    $0$      \\ \hline
$e_1$                 & ${\underline {\bf 1} } $       &    0
&
1      & \ldots             &  $0$                   &  $0$  \\ \hline
$\vdots $  &    $\vdots   $         &   $\vdots   $        &    $\vdots   $
&             $\ddots $     &     $\vdots   $        &     $\vdots   $
\\ \hline
$e_{n_c-r-1} $    &  ${\underline {\bf 1}} $    & 0                     & 0
&      $ \ldots  $            & 1                 &  0 \\ \hline
\end{tabular}
\caption{The effective degrees of freedom and their quantum numbers at the
``nonbaryonic root". }
\label{tabnonb}
\end{center}
\end{table}

Once the gauge group and  the quantum numbers of the matter fields are all known,  the ${\cal N}=2$  supersymmetry 
   uniquely fixes the structure of the effective action.    We find that 
   
   \begin{description}
  \item[(i)]   We see 
  the non-Abelian monopoles in action,  in the generic $r$  ($2\le  r \le  \frac{N_{f}}{2}$)  vacua.   See Table \ref {tabsun}   taken from \cite{CKM}.    They behave  perfectly as  point-like particles,  albeit in a dual, magnetic gauge system. 
 Upon  ${\cal N}=1 $ perturbation  they condense (confinement  phase)
  $ \brc  q_{a}^{i} \ckt  \sim   \delta_{a}^{i} \sqrt{\mu\, \Lambda}$     and 
  induces flavor symmetry breaking  
  \[   SU(N_{f}) \times U(1)  \to  U(r) \times U(N_{f}-r).
  \]
 
  \item[(ii)]   The upper limit  $ r \le  \frac{N_{f}}{2}$  is a manifestation of monopole dynamics:  only in this range of $r$   the non-Abelian monopoles can appear as {\it recognizable infrared degrees of freedom.}     We  now see  why   in the $SU(2)$ Seiberg-Witten models, as well as in pure ${\cal N}=2$ Yang-Mills  ({\i.e.}, $N_{f}=0$) models   with different gauge groups,   the low-energy  monopoles 
 were found to be  always Abelian:   in all these cases, non-Abelian monopoles would interact too strongly, not enough of them being there.  We remind the reader that the beta function in    ${\cal N}=2$  $SU$ theories has the pure one-loop form with  $\beta_{0} \propto  2 \, r -   N_{f}$.   
 
 \item[(iii)]    Indeed, there are  homotopy and  symmetry arguments   \cite{CKM,ABEKM} which suggest that  non-Abelian  monopoles appearing in the $r$-vacua are  ``baryonic constituents''
  of an Abelian  ('t Hooft-Polyakov) monopople, 
  \be     Abelian \,\, monopole \sim            \epsilon^{a_1   \ldots  a_r} {q_{a_1}^{i_1}  q_{a_2}^{i_2}
\ldots q_{a_r}^{i_r} },     
  \ee
  $a_{i}$ being the dual color indices, and $i_{m}$ the flavor indices.   The  $SU(r)$  gauge ineractions, being infrared-free, are unable to  keep  the Abelian monopole bound:    
they disintegrate   into non-Abelian  monopoles. 
 
  \item[(iv)]  That the effective degrees of freedom  in the $r$ vacua   are  non-Abelian rather than Abelian monopoles,  is actually required  also  by symmetry of the system \cite{CKM,MKY}, not only from the dynamics.  If the Abelian  monopoles  of the $r$-th tensor flavor  representation 
  were the correct degrees of freedom,  
     the low-energy effective theory would have too large an accidental symmetry --  $SU({N_{f} \choose r})$.        The condensation of such monopoles would produce far-too-many Nambu-Goldstone bosons than expected  from the symmetry of the underlying theory.   The system prevents  such an awkward situation from being  realized  in an elegant manner,  introducing smaller solitons, non-Abelian monopoles, in the fundamental representation of the $SU(N_{f})$     so that the low-energy theory has the right symmetry. 
  
\item [(v)]  An analogous  argument might be used  in  the standard QCD, to exclude Abelian picture of confinement, though admittedly this is not a very  rigorous one. 
We know from lattice simulations of $SU(3)$  theory  that confinement and chiral symmetry breaking are closely related.   If Abelian 't Hooft-Monopole-Mandelstam
monopoles were  the right degrees of  freedom describing confinement,  their condensation would  somehow have to describe chiral symmetry breaking as well.    We would then be  led to assume that they carry flavor quantum numbers of $SU(N_{f})_{L}\times SU(N_{f})_{R}$,   {\it e.g.}, 
\[    Monopoles \sim  M_{i}^{j}, \qquad   \brc    M_{i}^{j}  \ckt  \propto  \delta_{i}^{j} \, \Lambda_{QCD}, \]
 where $i, j$ are     $SU(N_{f})_{L}\times SU(N_{f})_{R}$ indices.   But such a system would have  a far-too large accidental symmetry.  Confinement would be accompanied by a large number of unexpected  (and indeed unobserved)   light Nambu-Goldstone bosons.
 
\item[(vi)]     The limiting case  of $r$ vacua, with  $r= \frac{N_{f}}{2}$, as well as the  massless   ($m_{i}
 \to  0$)  limit of   $USp(2N_{c})$ and $SO(N_{c})$   theories,  are of  great  interest.  The low-energy effective theory in these cases turn out to be 
 conformally invariant (nontrivial infrared-fixed-point) theories. This is an analogue of an Abelian superconformal vacuum   found  first in the pure $SU(3)$ Yang Mills theory 
 by Argyres and Douglas \cite{AD}.  It can be explicitly checked that the low-energy degrees of freedom include  relatively non-local  monopoles and dyons \cite{CKM,AGK,MKY}.   There are no  local effective Lagrangians describing the infrared dynamics. 
 These are the most difficult cases to analyze, 
 but are  potentially the most interesting ones,  from the point of view of understanding QCD.  
 We shall come back  to these (perhaps, crucial)  cases at  the end of the lecture, Section \ref{SCF}.     
\end{description}   
  
\begin{table}[h]

\begin{center}
  {\small   {
\begin{tabular}{ccccc}
    label ($r$)    &   Deg.Freed.      &  Eff. Gauge  Group
&   Phase    &   Global Symmetry     \\
\hline
$0$      &   monopoles   &   $U(1)^{n_c-1} $               &   Confinement
   &      $U(n_f) $            \\ \hline
$ 1$            &  monopoles         & $U(1)^{n_c-1} $        &
Confinement       &     $U(n_f-1) \times U(1) $        \\ \hline

$ \le  [\frac{n_f -1}{  2}] $    & NA monopoles      &    $SU(r)
\times U(1)^{n_c-r}   $  &    Confinement
&          $U(n_f-r) \times U(r) $
\\ \hline
$ {n_f / 2}  $  &   rel.  nonloc.     &    -    &    Confinement
&          $U({n_f / 2} ) \times U({n_f/2}) $          
\\ \hline
BR  &  NA monopoles     &
$ SU({\tilde n}_c) \times  U(1)^{n_c -  {\tilde n}_c } $                &
Free Magnetic
&      $U(n_f) $         \\ \hline
\end{tabular}  }}
\caption{ { Phases of $SU(n_c)$ gauge theory with $n_f$ flavors.       
$ {\tilde n}_c
\equiv n_f-n_c$. } }  
\label{tabsun}
\end{center}
\end{table}

\begin{table}[h]

\begin{center}
  {\small {
\begin{tabular}{ccccc}
        &   Deg.Freed.      &  Eff. Gauge Group
&   Phase    &   Global Symmetry     \\
\hline
1st Group  &  rel.  nonloc.       &    -    &
Confinement
&          $ U(n_f)  $
\\ \hline
2nd Group       &  dual quarks     &      $USp(2  {\tilde n}_c) \times
U(1)^{n_c -{\tilde n}_c} $               &  Free Magnetic
&      $SO(2n_f) $         \\ \hline
\end{tabular}
  }}
\caption{{ Phases of $USp(2 n_c)$ gauge theory  with $n_f$ flavors  with
$m_i \to 0$.   $ {\tilde n}_c \equiv n_f-n_c-2$. }}
\label{tabuspn}
\end{center}
\end{table} 
\begin{figure}
\begin{center}
\includegraphics[width=2.5 in]{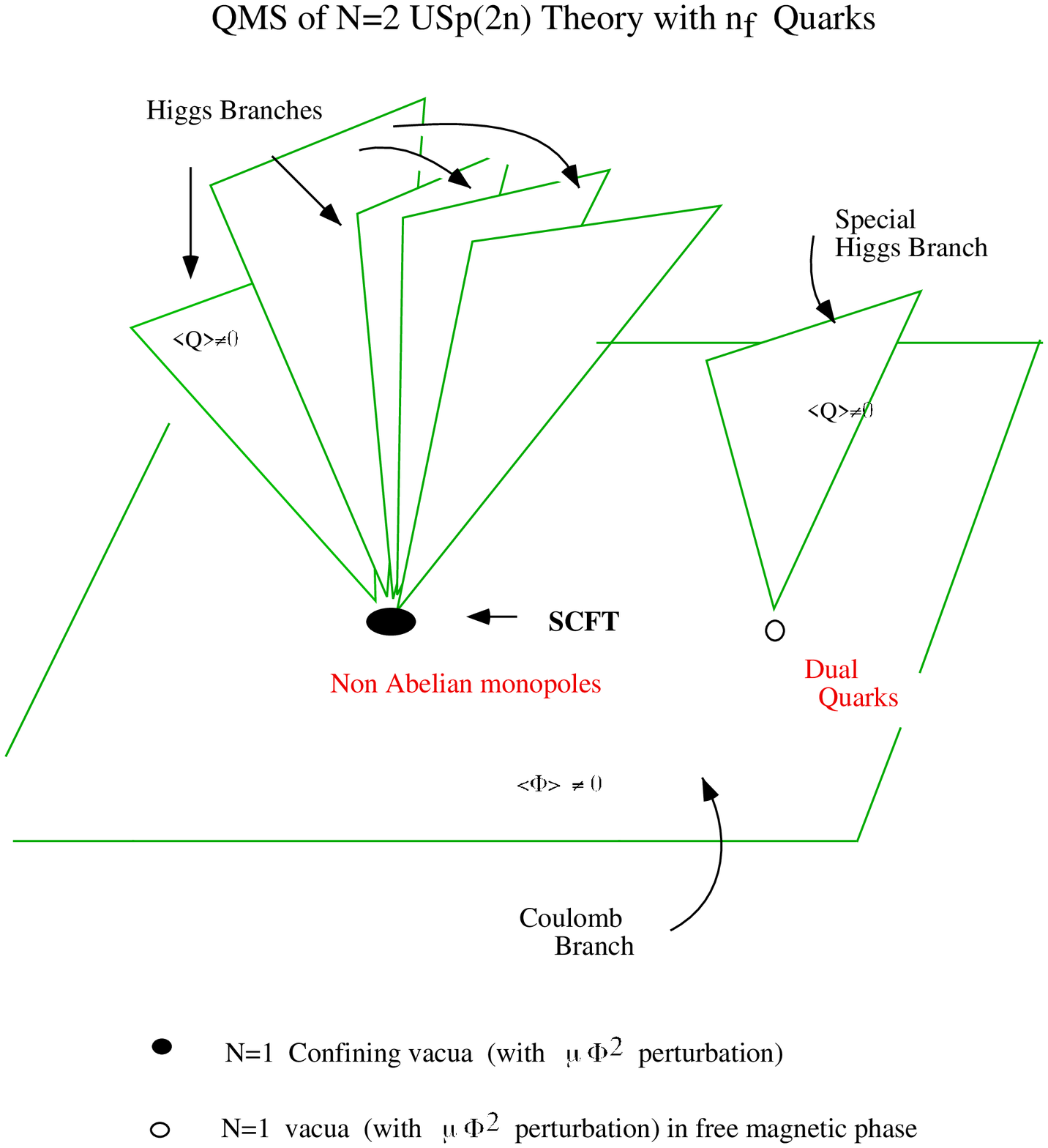}
\caption{ }
\label{UspQMS}
\end{center}
\end{figure}

\section{Vortices}  

The moral of the story is  that the non-Abelian monopoles do exist in fully quantum mechanical systems.  In typical  confining vacua in supersymmetric gauge theories  they are the relevant infrared degrees of freedom.   Their condensation induces confinement and dynamical symmetry breaking.     This brings us back to the problem  of {\it understanding}  these light, magnetic degrees of freedom as quantum solitons: 
\begin{center}
{\tt
  What are  their semi-classical counterparts? 
  
    Are they   Goddard-Nuyts-Olive-Weinberg monopoles?  
    
    In which sense condensation of non-Abelian monopoles imply confinement?  
    
     How has  the difficulty related to the dual group mentioned earlier  been avoided? }
\end{center} 
These are the questions we wish to answer.  The idea is to take advantage of the fact that in supersymmetric theories there are parameters which can 
be varied, upon which the physical properties of the system depend in a holomorphic fashion.   As $m_{i}$ and $\mu$ are varied, there cannot be phase transition at some $|\mu|$ or at $|m_{i}|$: the number of Nambu-Goldstone bosons and hence the pattern of the symmetry breaking,  must be invariant.

\subsection{Abrikosov-Nielsen-Olesen vortex} 

Topologically stable vortices arise when the ground states of a system have a nontrivial moduli space which is not simply connected.  
The best known case \cite{ANO}   is the Abelian gauge theory with a charged complex matter field  in Higgs phase (superconductor),  where the static configurations have energy density 
\[    H  =  \frac{1}{4}  F_{ij}^{2}  +    |D_{i} \phi |^{2}  +  V(|\phi|),  \qquad   D_{i}=  \de_{i} -  i \, e\, A_{i}.
\]
The potential  $V$  is assumed to attain its minimum  at $|\phi|=v \ne 0$.     The asymptotic gauge and scalar fields must be such that the field energy be finite, 
\[     |\phi(x)| \to  v, \qquad          D_{i} \phi \to 0,  \qquad F_{ij}^{2} \to 0.  
\]
These allow for nontrivial configurations   classified by an integer,  
\[  \pi_{1}(U(1)) ={\mathbf Z}, 
\]
{\it i.e.}, by an integer winding number $n$, 
\[     \phi \to   e^{i n \varphi}\, v,   \qquad   A_{\varphi} \to  \frac{n}{e \rho},   
\]
where $\rho, \phi, z$ are  the position variables of cylindrical coordinate system.  
At the center of the vortex  $\phi(\rho=0, \varphi) =0$  in order for $\phi(\rho, \phi)$ to be a smooth configuration: the gauge symmetry is restored along    the vortex core.   

Depending on the potential,  the vacuum can be superconductor of  type II where single isolated 
 (Abrikosov-Nielsen-Olesen) vortices  are  stable, type I systems where vortices stick together  to form the regions of normal ground state, and  finally there is the critical case  between them (BPS) where vortices has no net  interaction  and the tension of winding number $k$ vortex is equal to $k$ times that of the minimum-winding  vortex. 

\subsection{${\mathbf Z}_{N}$   vortices}

In pure $SU(N)$ theory with all matter fields in adjoint representation,  the true gauge group is  $\frac{SU(N)}{{\mathbf Z}_{N}}$. 
When the gauge group is completely broken the vacuum manifold has nontrivial structure, 
\be \pi_{1}(\frac{SU(N)}{{\mathbf Z}_{N}})=  {\mathbf Z}_{N}.\label{homoZN}
\ee
 The asymptotic behavior of the fields, required by finiteness of the tension is 
$$   A_i  \sim   \frac{ i }{  g}    \, U(\phi) \de_i  U^{\dagger}(\phi);    
\quad     \phi_A \sim   U \phi_{A}^{(0)}  U^{\dagger},   \qquad  U(\phi) = \exp{i \sum_j^r 
\beta_j T_j
\phi}
$$
where $T_j$  are the generators of the Cartan subalgebra of $H$, $ \phi_{A}^{(0)}$ are the (set of)  VEVs of the adjoint scalar fields 
which  break the $SU(N)$ group completely.   
 The smoothness of the configurations requires the quantization condition:  
($\alpha$ =  root vectors of $H$) 
\be   U(2 \pi)  \in    {\bf Z}_N,  
\qquad   
{ \alpha} \cdot {  \beta}  \in  {\bf Z}.      
\label{ZN}\ee
The second condition of (\ref{ZN})  appears to imply that these vortices be characterized by 
the {\it weight vectors} of  the group ${\tilde H}=SU(N)$, dual of $H=SU(N)/{\bf Z}_N$ \cite{GNO}:   one vortex for each irreducible representation of ${\tilde H}$. 
Actually, Eq. (\ref{homoZN})  shows that there is just one stable vortex  with a given ${\bf Z}_N$ charge ($N$-ality). \footnote{That an excitation in a theory in which all fields are neutral with respect to ${\mathbf Z}_{N}$   is characterized by a fractional  ${\bf Z}_N$ charge, may be thought of as an analogue of a very general behavior of solitons:  charge fractionalization.\footnote{GoldWil} }     

An interesting  model of this sort is the so-called $N=1^{*}$ theory \cite{WittN1,StrassLat,HSZ} defined as the ${\cal N}=4$ supersymmetric theory with addition of mass terms for the three adjoint scalar multiplets, 
\[    \Delta L=   \sum_{i=1}^{3}  m_{i}\,  \Phi_{i}^{2}|_{\theta \theta},
\]
which break supersymmetry to ${\cal N}=1$.  The general properties  of chiral condensates, 
\[     \brc  W\, W \ckt, \quad  \brc \Phi_{1}^{2} \ckt, \quad  \brc \Phi_{2}^{2} \ckt,  \quad  \brc \Phi_{3}^{2} \ckt, 
\]
in all possible types of vacua  (confinement vacua, Coulomb vacua, Higgs vacua) have been analyzed exactly in a series of papers \cite{Nick}. 

This model is based on the underlying ${\cal N}=4$ model, which is believed to display exact Olive-Montonen duality. 
In spite of the  relative simplicity of the model, the properties of ${\bf Z}_N$ monopoles in the Higgs (or partially Higgs) vacua in the ${\cal N}=1^{*}$  are not very well known, except for the $SU(2)$  \cite{MMY}   or $SU(3)$  cases.         

\subsection{Non-Abelian vortices in a $U(N)$  model  \label{UNmodel}}  

The ${\bf Z}_N$  vortice discussed in the preceding section at first sight  appears to carry a non-Abelian charge,  being labelled by the weight vector of a non-Abelian dual group   ${\tilde H}$:  actually,  they do not \cite{KS}.  It is just a single solution, which can be transformed by Weyl transformations of $H$.  There are no continuous moduli associated to it. 

Truly non-Abelian  vortices have been constructed \cite{HT,ABEKY}   in the context of a ${\cal N}=2$   supersymmetric $U(N)$  gauge theory, with  $N_{f} $ flavors, where the gauge group is broken by  the VEVs   of a set of scalar fields  in the fundamental representations.
The model Lagrangian has the form
\bqa 
{\cal L} &=& \Tr \left[
- \frac{1}{2g^2} F_{\mu\nu}F^{\mu\nu}- \frac{2}{ g^2}  \D_\mu \,\phi^{\dagger} \,\D^\mu  \phi - \D_\mu \,H \,\D^\mu  H^\dagger
- \lambda \left( c\,{\bf 1}_N - H\,H^\dagger\right)^2 \right] \nonumber  \\
&+&   \Tr \, [ \, (H^{\dagger} \phi  - M \,  H^{\dagger} )
(  \phi  \, H -  H \, M ) \,]  \label{widely}
\eea
where $F_{\mu\nu} = \partial_\mu W_\nu - \partial_\nu W_\nu + i \left[W_\mu,W_\nu\right]$
and $\D_\mu H  = \left(\partial_\mu + i\, W_\mu\right)\, H$,
and $H$  represents  the  fields in the fundamental representation of $SU(N)$,    written in  a color-flavor $N \times N_{f}$  matrix form,   $(H)_{\alpha}^{i} \equiv q_{\alpha}^{i}  $,  and $M$ is a $N_{f}\times N_{f}$ mass matrix.
Here, $g$ is the $U(N)_{\rm G}$ gauge coupling,  $\lambda$ is a
scalar coupling. For
\be   \lambda = \frac{g^2}{4}\ee
  the system is BPS  saturated.
 For such a choice,  the   Eq. (\ref{widely})   can be regarded as a  truncation
  of  the bosonic sector of an  ${\cal N}=2$ supersymmetric $U(N)$ gauge theory, and with $ (H)_{\alpha}^{i} $ representing the half of the squark fields,
   \be    (H)_{\alpha}^{i} \equiv q_{\alpha}^{i}, \qquad {\tilde q}^{\alpha}_{i}  \equiv 0   \ee
   In the supersymmetric context the  parameter  $c$ is   the Fayet-Iliopoulos
parameter. In the following we set $c>0$  so that the system be in Higgs phase, and so as to allow stable vortex configurations.
For generic, unequal quark masses,
\be     M =  diag \, (m_{1},m_{2},\ldots, m_{N_{f}}),   \label{unequalmcase}
\ee
the adjoint scalar VEV takes the form,
\be    \brc \phi \ckt  =  M =   \left(\begin{array}{cccc}m_1 & 0 & 0 & 0 \\0 & m_2 & 0 & 0 \\0 & 0 & \ddots & 0 \\0 & 0 & 0 & m_N\end{array}\right),  \label{unequalmphi}  \ee
which breaks the gauge group to $U(1)^{N}$.

In  order to have a non-Abelian  vortex,  it is  necessary to choose masses equal,
\be   M =  diag \,  (m,m,\ldots, m),   \label{equalmcase}
\ee
the adjoint and squark fields have the vacuum expectation value (VEV)
\be    \brc \phi \ckt  =  m \, {\mathbf 1}_N,  \qquad \brc H  \ckt = \sqrt{ c} \,  \left(\begin{array}{ccc}1 & 0 & 0 \\0 & \ddots & 0  \\0 & 0 & 1   \end{array}\right) \label{symmbr}
\ee
where only the first $N$ flavors are left explicit.  
The squark VEV breaks the gauge symmetry completely, while leaving an unbroken $SU(N)_{C+F}$ color-flavor diagonal symmetry (the flavor group acts on $H$ from the right
while the $U(N)_{\rm G}$  gauge symmetry acts on $H$ from the left).  The global symmetry group associate with the other $N_{f}-N$  flavors also remains unbroken.  The BPS vortex equations are
\be 
\left(\D_1+i\D_2\right) \, H = 0,\quad
F_{12} + \frac{g^2}{2} \left( c \,{\bf 1}_N - H\, H^\dagger\right) =0.
\ee
The matter  equation can be solved \cite{Isozumi:2004vg}-\cite{Eto:2006pg}   by use of the $N \times N$  moduli matrix $H_0(z)$
whose components are holomorphic functions of the  complex coordinate $z = x^1+ix^2$,
\be 
H = S^{-1}(z,\bar z) \, H_0(z),\quad
W_1 + i\,W_2 = - 2\,i\,S^{-1}(z,\bar z) \, \bar\partial_z S(z,\bar z).
\ee
The gauge field equations then take the simple form  (``master equation'')  
\be 
\de_{z}\,(\Omega^{-1} \de_{\bar z} \, \Omega ) =  \frac{g^{2}}{4} \, ( c\, {\mathbf 1}_N -  \Omega^{-1} \, H_{0}\, H_{0}^{\dagger} ).
   \label{master} \ee
The  moduli matrix and $S$ are defined up to a  redefinition,
\be    H_{0}(z)   \to  V(z) \, H_{0}(z), \qquad S(z, \bar{z})   \to  V(z) \, S(z, \bar{z}), \label{redef}
\ee
where  $V(z)$ is any non-singular  $N \times N$  matrix which is holomorphic in $z$.
This class of model has been extensively studied recently \cite{Isozumi:2004vg}-\cite{SYSemi}.  
In particular,  in the contex of these models a considerable attention was given to the system in which 
$U(N)$ gauge symmetry is either explicitly or dynamically broken to $U(1)^{N}$, producing {\it Abelian}  monopoles.    As the  
terminology used and concepts  involved, though physically distinct,  are often similar to the concept of non-Abelian monopoles discussed in this note, and could be misleading.

\subsection{ Dynamical Abelianization \label{sec:abelian} }

\begin{figure}
\begin{center}
\includegraphics[width=3in]{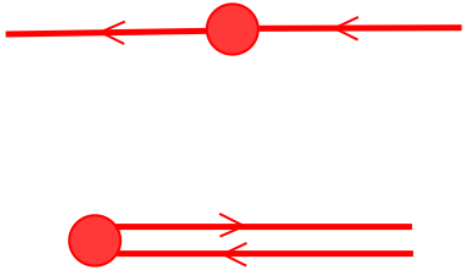}
\caption{Monopoles in  $U(N)$   systems with Abelianization are confined by two Abelian  vortices. }
\label{two}
\end{center}
\end{figure}

As should be clear from what we said so far, it is crucial that the color-flavor diagonal symmetry $SU(N)$ remains  exactly conserved, for the emergence  of non-Abelian dual gauge group (see the next Section).  Consider, instead,  the cases in which  the gauge $U(N)$  (or $SU(N) \times U(1)$)  symmetry is broken to   Abelian subgroup  $U(1)^{N}$,  either by small quark mass differences   (Eq. (\ref{unequalmphi}))  or  dynamically,  as in the ${\cal N}=2$  models  with  $N_{f} <  2\, N$ \cite{HT2,SY}.     From the breaking of  various $SU(2)$ subgroups  to $U(1)$ there appear light 't Hooft-Polyakov  monopoles of mass  $O(\frac{\Delta m}{g})$ (in the case of an explicit breaking)  or   $O(\Lambda)$  (in the case of dynamical breaking).   As the  $U(1)^{N}$ gauge group is further broken by the squark VEVs,  the system develops ANO vortices.  The light magnetic monopoles,  carrying  magnetic  charges  of two  different  $U(1)$ factors,  look  confined  by the two vortices  (Fig. \ref{two}).    These cases have been discussed extensively \cite{Eto:2006pg}-\cite{GSY}, within the context of $U(N)$  model of  Subsection \ref{UNmodel}.

  The dynamics of the fluctuation of  the orientational modes along the vortex  turns out to be  described by a two-dimensional  ${\bf C}P^{N-1}$ model \cite{HT,ABEKY}.   It has been  shown \cite{HT,HT2,SY,GSY}, that the kinks of the two-dimensional sigma model precisely correspond to these light monopoles, to be expected in the underlying $4D$ gauge theory.  In particular, it was noted that there is an elegant matching  between the dynamics of two-dimensional sigma model (describing the dynamics of the  vortex orientational modes in the Higgs phase of the $4D$ theory)   and the dynamics of the $4D$ gauge theory in the Coulomb phase, including the precise matching of the 
  coupling constant renormalization \cite{Tong,HT2,SY}.

Note that these cases are analogue of  {\it what would occur} in QCD if the color $SU(3)$  symmetry were to dynamically break itself  to $U(1)^{2}$.     Confinement would be described in this case by the condensation of  magnetic monopoles carrying the Abelian charges $Q^{1}$, or $Q^{2}$, and the resulting ANO vortices will be of two types, $1$ and $2$  carrying the related fluxes.

\section{The  model}

Actually the model we need here is not exactly the model of Section \ref{UNmodel}, but is a model which contains it  as a low-energy 
approximation.    It is the same model already discussed in Section \ref{withquarks}, but now we analyze it in the region, $m_{i} \gg
\mu \gg \Lambda$, so that  the semiclassical reasoning of Section \ref{relating}
makes sense.  For concreteness, 
we take as our model the standard ${\cal N}=2$ SQCD with  $N_{f}$ quark hypermultiplets,   with a larger
 gauge symmetry, {\it e.g.},  $SU(N+1)$,   which is broken at a much larger mass scale ($v_{1}   \sim |m_{i}|$)     as
\be    SU(N+1)  \,\,\,{\stackrel  {v_{1}    \ne 0} {\longrightarrow}} \,\,\, \frac{ SU(N) \times U(1)}{{\mathbf Z}_{N}}.
\ee
The unbroken gauge symmetry is completely broken at a lower mass scale, 
$v_{2}\sim |\sqrt{\mu  m}|$,      as in Eq. (\ref{squarkvev}) below.

Clearly one can attempt  a similar embedding of the  model Eq. (\ref{widely})  in a larger gauge group broken at some higher
 mass scale, in the context of a non-supersymmetric model,  even though in such a case the potential must be judiciously chosen
  and the dynamical  stability of the scenario would have to be carefully monitored.   Here we choose to study the softly broken 
  ${\cal N}=2$ SQCD  for  concreteness,  and above all because  the dynamical properties of this model are well understood:  this will
   provide us with a non-trivial check of our results.  Another motivation is purely of convenience: it gives a definite potential with 
    desired properties.\footnote{
Recent developments \cite{DV,CDSW} allow us actually  to consider systems of this sort  within
 a much wider class of ${\cal N}=1$ supersymmetric models, whose infrared properties are very much under control.  
}

We are hereby  back to our argument on the duality and non-Abelian monopoles,  defined through a better-understood 
non-Abelian {\it vortices} presented in general terms in 
Section \ref{sec:nonamonopoles},  but now in the context of a concrete model, where the fully quantum mechanical 
answer is known.

The underlying theory is thus
\be   {\cal L}=     \frac{1}{ 8 \pi} {\Im} \, S_{cl} \left[ \int d^4 \theta \,
\Phi^{\dagger} e^V \Phi +\int d^2 \theta\,\frac {1}{ 2} W W\right]
+ {\cal L}^{({\rm quarks})}  +  \int \, d^2 \theta \,\mu  \,\Tr  \Phi^2 + h.c. ;
\label{lagrangian}
\ee
\bqa && {\cal L}^{({\rm quarks})}   =  \\
&&  \sum_i \, \left[ \int d^4 \theta \, \{ Q_i^{\dagger} e^V
Q_i + {\tilde Q_i}  e^{-V} {\tilde Q}_i^{\dagger} \} +   \int d^2 \theta
\, \{ \sqrt{2} {\tilde Q}_i \Phi Q^i    +      m_{i} \,    {\tilde Q}_i \, Q^i   \}+ h.c. \right]   \non 
\eea
where $m_{i}$ are  the bare masses  of the quarks and we have defined the complex coupling constant
\be
S_{cl} \equiv  \frac{\theta_0}{\pi} +\frac {8 \pi i }{ g_0^2}.
\ee
We also added the  parameter $\mu$, the mass of the adjoint chiral multiplet, which breaks the supersymmetry softly to ${\cal N}=1$.
The bosonic sector of this model  is described, after elimination of  the auxiliary fields, by
\be  {\cal L}=  \frac{ 1 }{4 g^2}  F_{\mu \nu}^2  +  \frac { 1}{ g^2}  |{\cal D}_{\mu} \Phi|^2 +
 \left|{\cal D}_{\mu}
Q\right|^2 + \left|{\cal D}_{\mu} \bar{\tilde{Q}}\right|^2 -    V_1 -  V_2,
\label{Lag}\ee
where
\be    V_1   =          \frac { 1}{ 8 }  \, \sum_A  \left( t^A_{ij} \,[ \,  \frac { 1}{ g^2 }  (-2)  \, [\Phi^{\dagger},   \Phi]_{ji}  +  Q^{\dagger}_j   Q_i -  {\tilde Q}_j   {\tilde
Q}^{\dagger}_i \,] \right)^2;
\ee
\bqa  V_2 &=&      g^2 |   \mu \, \Phi^A +
\sqrt 2   \, {\tilde Q} \, t^A  Q |^2    +     {\tilde Q } \,   [ m    + \sqrt2  \Phi   ] \,  [ m    + \sqrt2  \Phi   ]^{\dagger}  \, {\tilde Q}^{\dagger}
\nonumber  \\  & +  & Q^{\dagger} \, [ m    + \sqrt2  \Phi   ]^{\dagger}    \, [ m    + \sqrt2  \Phi   ]  \, Q.
\eea
In the construction of the approximate monopole and vortex  solutions  we shall consider only the VEVs and fluctuations around them which satisfy
\be   [\Phi^{\dagger},   \Phi]=0, \qquad    Q_i   =  {\tilde Q}^{\dagger}_i,  \label{trunc1}
\ee
and hence  the $D$-term potential  $V_1$   can be set identically to zero throughout.

In order to keep the hierarchy of the gauge symmetry breaking scales, Eq. (\ref{hierarchybis}), we
choose the masses such that
\be     m_{1}=\ldots = m_{N_{f}}= m, \label{equalmass}  \ee
\be    m   \gg  \mu    \gg   \Lambda.  \label{doublescale}
\ee
Although the theory described by the above Lagrangian has many degenerate vacua, we are interested in the vacuum where\ldots
(see \cite{CKM} for the detail)
\be    \brc\Phi  \ckt =  - \frac{1}{\sqrt 2}  \, \left(\begin{array}{cccc} m & 0 & 0 & 0 \\0 & \ddots & \vdots  & \vdots \\0 & \ldots   & m & 0 \\0 & \ldots & 0 & - N \, m \end{array}\right);   \label{adjointvev}
\ee
\be
Q=  {\tilde Q}^{\dagger}  = \left(\begin{array}{ccccc}  d & 0 & 0 &  0 & \ldots \\0 & \ddots  & 0 & \vdots & \ldots \\0 & 0 & d &  0 & \ldots \\0 & \ldots & 0 &  0 & \ldots\end{array}\right),  \qquad   d = \sqrt{ (N+1)\, \mu \, m}. \label{squarkvev} \ee
This is a particular case of the so-called $r$ vacuum, with $r=N$.  Although such a vacuum certainly exists classically, the existence of the  quantum $r=N$  vacuum in this theory requires  $N_{f} \ge  2\, N$,
which we shall assume.\footnote{
This might appear to be a rather tight condition as the original theory loses asymptotic freedom  for  $N_{f} \ge  2\, N + 2$.  This is not so.  An analogous discussion can be made by  considering the breaking
$SU(N) \to SU(r) \times U(1)^{N-r}$.   In this case  the condition for the quantum non-Abelian  vacuum is  $2\, N > N_{f} \ge  2\, r$, which  is a much looser condition.   }


%

  To start with, ignore the smaller squark VEV, Eq. (\ref{squarkvev}). As $\pi_{2}(G/H) \sim \pi_{1}(H) = \pi_{1}(U(1)) ={\mathbf Z}$,
 the symmetry breaking  Eq. (\ref{adjointvev})  gives rise to regular magnetic monopoles with mass of order of $O(\frac{v_{1}}{g})$,  whose continuous transformation property is our main concern here.

 The semiclassical formulas for their mass and fluxes  \cite{EW,ABEKM} are summarized in Appendix  \ref{sec:General}.

\subsection {Low-energy approximation and vortices}

At scales much lower than  $v_{1} = m$ but still neglecting the smaller squark  VEV
$v_{2} =   d = \sqrt{ (N+1)\, \mu \, m} \ll  v_{1}$,    the theory reduces to an $SU(N)\times U(1)$  gauge theory with $N_{f}$ light quarks $q_{i}, {\tilde q}^{i}$ (the first $N$ components of the original quark multiplets $Q_{i}, {\tilde Q}^{i}$).  By integrating out the massive fields, the effective Lagrangian valid between the two mass scales has the form,
\bqa  {\cal L} &=&  \frac{ 1}{ 4 g_N^2}  (F_{\mu \nu}^a)^2  + \frac{ 1}{ 4 g_1^2}  (F_{\mu \nu}^0)^2  +  \frac{ 1}{ g_N^2}  |{\cal D}_{\mu} \phi^a |^2 +
 \frac{ 1}{ g_1^2}  |{\cal D}_{\mu} \phi^0 |^2 +
 \left|{\cal D}_{\mu}
q \right|^2 + \left|{\cal D}_{\mu} \bar{\tilde{q}}\right|^2  \nonumber \\
&-&    g_1^2 \bigg|  -  \mu \, m  \sqrt{N(N+1)}  +
\frac{ {\tilde q} \,  q}{\sqrt{N(N+1)}}    \, \bigg|^2
 - g_N^2 |\sqrt 2   \, {\tilde q} \, t^a q \,  |^2  + \ldots
\label{leappr}    \eea
where $a=1,2,\ldots N^{2}-1$ labels the $SU(N)$ generators,  $t^a$;   the index $0$ refers to the $U(1)$ generator $t^0= \frac { 1 }{ \sqrt {2N(N+1)}} \, diag  (1, \ldots, 1, - N).$  We have taken into  account the fact that  the $SU(N)$ and $U(1)$   coupling constants ($g_N$ and   $g_1$) get renormalized differently towards the infrared.

 The adjoint scalars are fixed to its VEV,  Eq. (\ref{adjointvev}), with small fluctuations around it,
\be  \Phi =  \brc\Phi  \ckt  (1 +    \brc\Phi  \ckt^{-1} \, {\tilde  \Phi} )  , \qquad   |{\tilde  \Phi}| \ll m.
\label{small}\ee
In the consideration of the vortices of the low-energy theory,  they will be in fact replaced by the constant VEV.  The presence of the small terms Eq. (\ref{small}), however, makes the low-energy vortices not strictly BPS  (and this will be  important in the consideration of their stability below).\footnote{In the terminology used in Davis et al. \cite{Davis}  in the discussion of the Abelian vortices in supersymmetric models, our model corresponds to an F model while the models of \cite{Tong,SY,Etou} correspond to a D model. In the approximation of replacing $\Phi$ with a constant,  the two models are equivalent: they are related by an $SU_{R}(2)$ transformation \cite{VainYung,ABE}.
}

The quark fields are replaced,  consistently with Eq. (\ref{trunc1}),  as
\be    {\tilde q} \equiv   q^{\dagger}, \qquad   q \to  \frac{1}{\sqrt{2}} \, q,
\ee
where the second replacement brings back the kinetic term to the standard form.

We further replace  the singlet coupling constant and the $U(1)$  gauge field as
\be     e \equiv  \frac{g_{1}}{\sqrt{2 N(N+1)}};     \qquad   {\tilde A}_{\mu} \equiv   \frac{A_{\mu}}{\sqrt{2 N(N+1)}},  \qquad  {\tilde \phi}^{0}\equiv   \frac{\phi^{0}}{\sqrt{2 N(N+1)}}.
\ee
The net effect is
\be  {\cal L} =  \frac{ 1}{ 4 g_N^2}  (F_{\mu \nu}^a)^2  + \frac{ 1}{ 4 e^2}  ({\tilde F}_{\mu \nu})^2  +
 \left|{\cal D}_{\mu}
q \right|^2
-    \frac{e^2}{2} \, |
 \, q^{\dagger} \,  q   -     c \, {\mathbf 1}  \, |^2 - \frac{1}{2} \, g_N^2 \,| \,
 \,  q^{\dagger} \, t^a q \,  |^2,
\ee
\be  c=   N(N+1)  \sqrt {    2\, {  \mu \, m}    }  .
\ee
Neglecting the small  terms left implicit, this  is  identical to the $U(N)$  model Eq. (\ref{widely}), except for the fact that $e \ne g_{N} $ here.   The
transformation property of the vortices can be determined  from the moduli matrix, as was done in \cite{seven}.
Indeed, the system possesses BPS saturated vortices described by the linearized equations
\be
\left(\D_1+i\D_2\right) \, q = 0,
\ee
\be
F_{12}^{(0)} + \frac{e^2}{2} \left( c \,{\bf 1}_N - q\, q^\dagger \right) =0; \qquad F_{12}^{(a)} + \frac{g_{N}^2}{2}\, q_{i}^\dagger  \, t^{a}\, q_{i}  =0.
\ee
The matter equation can be solved exactly as in
\cite{Isozumi:2004vg,Etou,Eto:2006pg}  ($z = x^1+ix^2$) by setting
\be
q  = S^{-1}(z,\bar z) \, H_0(z),\quad
A_1 + i\,A_2 = - 2\,i\,S^{-1}(z,\bar z) \, \bar\partial_z S(z,\bar z),
\ee
where $S$ is an  $N \times N$ invertible matrix  over whole of the $z$ plane, and  $H_{0}$ is  the  moduli matrix, holomorphic in $z$.

The gauge field equations take a slightly more complicated form than in the $U(N)$ model  Eq. (\ref{widely}):
\be  \de_{z}\,(\Omega^{-1}\de_{\bar z} \, \Omega ) =  -  \frac{g_{N}^{2}}{2}\, \Tr \, (\,t^{a} \, \Omega^{-1} \, q \, q^{\dagger} )\, t^{a} -
\frac{e^{2}}{4 N}\, \Tr \, (\,  \Omega^{-1} q \, q^{\dagger} - {\mathbf 1}), \quad  \Omega =  S\, S^{\dagger}.
 \ee
The last equation reduces to the master equation  Eq. (\ref{master}) in the  $U(N)$  limit, $g_{N}=e.$

The advantage of the moduli matrix formalism is that all the moduli parameters appear in the holomorphic, moduli matrix $H_{0}(z)$.   Especially, the transformation property of the vortices under the color-flavor diagonal group can be studied by studying the behavior of the moduli matrix.  

\subsection {Dual gauge transformation from the vortex moduli \label{sec:duality}}

The concepts such as the low-energy BPS vortices or the  high-energy BPS monopole solutions   are  thus only approximate:
their explicit forms are valid only in the lowest-order approximation, in the respective kinematical regions.   Nevertheless,  there is a property of the system   which  is exact   and does not depend on  any approximation: the full system has an exact, global $SU(N)_{C+F}$ symmetry,
which is neither broken by the interactions nor by both sets of  VEVs,  $v_{1}$ and  $v_{2}$.   This symmetry is    broken by individual soliton vortex,  endowing the latter  with  non-Abelian orientational moduli, analogous to the  translational zero-modes of a kink.
Note that the vortex breaks the color-flavor symmetry as
\be   SU(N)_{C+F} \to  SU(N-1) \times U(1),
\ee
leading  to the moduli space of the minimum vortices  which is
\be 
 {\cal M} \simeq  {\bf C}P^{N-1} = \frac{SU(N) }{ SU(N-1) \times U(1)}.
\ee
The fact that this moduli coincides with the moduli of the quantum states of an  $N$-state quantum mechanical system,  is a first hint that the monopoles appearing at the endpoint of a vortex, transform as a fundamental multiplet ${\underline N}$ of a group $SU(N)$.

The moduli space of the vortices is described by the moduli matrix  (we consider here the vortices of minimal winding,  $k=1$)
\be   H_{0}(z)   \simeq  \left(\begin{array}{cccc}   1 & 0 & 0 & -a_1 \\   0 & \ddots & 0 & \vdots \\  0 & 0 & 1 & - a_{N-1} \\   0 & \ldots  & 0 & z\end{array}\right),
\label{minSUN}\ee
where the constants $a_{i}$, $i=1,2,\ldots, N-1$ are the coordinates of ${\bf C}P^{N-1}$.
Under $SU(N)_{C+F}$ transformation,  the squark fields transform as
\be     q \to U^{-1} \, q \,U,
\ee
but as the moduli matrix is defined  {\it modulo}  holomorphic redefinition Eq. (\ref{redef}), it is sufficient to consider
\be     H_{0}(z)  \to    H_{0}(z)  \, U.
\ee
Now, for an infinitesimal $SU(N)$ transformation  acting on a matrix of the form Eq. (\ref{minSUN}),  $U$ can be taken in the form,
\be     U = {\bf 1} + X, \qquad X = \left(\begin{array}{cc}{\bf 0 }  & {\vec \xi} \\  -  ({\vec \xi})^{\dagger} & 0\end{array}\right),
\ee
where  ${\vec \xi}$ is a small $N-1$ component  constant vector.  Computing  $H_{0} \, X$ and  making a $V$ transformation from the left to bring back
$H_{0}$ to the original form,  we find
\be     \delta a_{i} =   - \xi_{i}  -   a_{i} \, ({\vec \xi})^{\dagger}\cdot {\vec a},  \label{inhomo}
\ee
which shows that $a_{i}$'s indeed transform as the inhomogeneous coordinates of ${\bf C}P^{N-1}$.  In other words, the vortex represented by the moduli matrix  Eq. (\ref{minSUN})  transforms as a fundamental multiplet of $SU(N)$.\footnote{
Note that,  if a ${\underline N}$  vector
${\vec c} $ transforms as ${\vec c}  \to  ({\bf 1} + X)\, {\vec c}  $,    the inhomogeneous coordinates   $a_{i} =  c_{i}/c_{N}$
transform as in Eq. (\ref{inhomo}).
}

As an illustration consider the simplest case of $SU(2)$  theory.
In this case the  moduli matrix is simply \cite{Eto:2004rz}
\be    H_{0}^{(1,0)} \simeq \left(\begin{array}{cc}z-z_0 & 0 \\  -b_{0}  & 1\end{array}\right); \qquad
 H_{0}^{(0,1)}  \simeq \left(\begin{array}{cc} 1  &  - a_{0}   \\ 0 & z-z_0\end{array}\right).
\label{minimum}\ee
with the transition function between the two  patches:
\be b_{0}= \frac{1}{a_{0}}.  \label{simple}  \ee
The points on this ${\bf C}P^{1}$ represent all possible $k=1$ vortices.    Note that  points on the space of a quantum mechanical  two-state system,
 \be  \ket {\Psi} = a_{1} \ket{\psi_{1}} + a_{2}\, \ket {\psi_{2}},  \label{twostate} \qquad     (a_{1}, a_{2}) \sim  \lambda \, (a_{1}, a_{2}), \quad \lambda \in {\bf C},
 \ee
 can be  put in one-to-one correspondence with  the inhomogeneous coordinate of a  ${\bf C}P^{1}$,
\be      a_{0} =  \frac {a_{1}}{a_{2}}, \qquad    b_{0} =  \frac {a_{2}}{a_{1}}.  \label{correspondence}
\ee
In order to make this correspondence manifest, note that the minimal vortex Eq. (\ref{minimum})  transforms under the $SU(2)_{C+F}$ transformation, as
\be   H_{0} \to   V  \, H_{0}  \, U^{\dagger}, \qquad  U= \left(\begin{array}{cc}\alpha & \beta \\-\beta^* & \alpha^{*} \end{array}\right), \quad
|\alpha|^{2} + |\beta|^{2} =1,
\ee
where the factor $ U^{\dagger}$ from  the right represents a flavor transformation,
$V$ is a holomorphic matrix  which brings $H_{0}$ to the original triangular  form \cite{seven}.  The  action of this transformation on the moduli parameter, for instance, $a_{0}$,  can be found to be
\be     a_{0} \to   \frac {\alpha \, a_{0} + \beta }{\alpha^{*} -  \beta^{*} \, a_{0}}.
\ee
But  this is precisely the way a doublet state  Eq. (\ref{twostate})
  transforms under $SU(2)$,
\be   \left(\begin{array}{c}a_1 \\a_2\end{array}\right)  \to    \left(\begin{array}{cc}\alpha & \beta \\-\beta^* & \alpha^{*} \end{array}\right) \, \left(\begin{array}{c}a_1 \\a_2\end{array}\right).
\ee

The fact that the  vortices (seen as solitons of the low-energy approximation)   transform as in the ${\underline N}$ representation of $SU(N)_{C+F}$, implies that
 there exist a set of monopoles  which transform accordingly, as   ${\underline N}$.  The existence of such a set follows from the exact $SU(N)_{C+F}$ symmetry of the theory, broken by the
individual   monopole-vortex configuration.  

This answers some of the questions formulated earlier (below  Eq. (\ref{question})) unambiguously \cite{seven}.
Note that in our derivation of continuous transformations of the monopoles,  the explicit, semiclassical form of the latter is not used.

A subtle point is that  in the high-energy approximation, and to lowest order of such an approximation,   the semiclassical monopoles are just certain non-trivial field configurations
involving $\phi(x)$ and $A_{i}(x)$ fields only, and  therefore apparently  transform under the color part of  $SU(N)_{C+F}$  only.  When the full
monopole-vortex configuration  $\phi(x), A_{i}(x), q(x) $  (Fig. \ref{monovortex}) is  considered,  however,  only the combined color-flavor diagonal transformations keep  the energy of the configuration invariant.   In other words,  the monopole transformations must be regarded as  part of more complicated transformations involving  flavor,   when  higher order  effects  in $O(\frac{v_1}{v_2})$   are taken into account.
And this means that the transformations are among {\it physically distinct states},   as the vortex moduli describe obviously physically distinct
vortices \cite{ABEKY}.

This discussion highlights the crucial role played by the (massless)  flavors  in the underlying theory   as has been already summarized at the end of Section 2.    There is, however, 
another  important  independent effect due to the massless flavors. Due to the zero-modes of the fermions, the semi-classical monopoles are converted to some irreducible  multiplets   in  the {\it flavor}  group $SU(N_{f})$ \cite{JR}. The ``clouds'' of the fermion zero-mode fluctuation fields surrounding the monopole have an  extension of  $O(\frac{1}{v_{1}})$, which is much smaller than the distance scales associated with the infrared effects discussed here.  We conclude that there was one more crucial  role of the flavor on non-Abelian monopoles:  it allows to generate the dual magnetic gauge group on the one hand,    and to ``dress''   the monopoles  and endow them with global, flavor quantum numbers \`a la Jackiw-Rebbi, on the other.  They should be regarded as two, distinct effects.

Our construction  has been generalized to the symmetry breaking $SO(2N+1) \to  U(N) \to {\bf 1}$, 
$SO(2N+1) \to  U(r) \times  U(1)^{N-r}  \to {\bf 1}$,   in the concrete context of  
softly broken ${\cal N}=2$   models.   There is an interesting difference in the quantum fate of the semiclassical monopoles in the case the unbroken $SU$ factor has the maximum rank and in the cases where $r \le  N-1$. The semiclassical (vortex-monopole complex) argument  of Section \ref{relating} and in this Section   and the fully quantum mechanical results  (of Section \ref{withquarks},  Section \ref{sec:exactqb}) agree qualitatively, quite nontrivially \cite{seven}.

The fact that the vortices of the low-energy theory are   BPS saturated,  which allows us to analyze their moduli and transformation properties elegantly  as discussed above, while in the full theory there are corrections which make them non BPS (and unstable), might cause some concern.  Actually, the rigor of our argument is not affected by those terms which can be treated as perturbation.  The attributes  characterized by integers such as the transformation property of certain configurations as a multiplet of a  non-Abelian group which is an exact symmetry group of the full theory,  cannot receive renormalization. This is similar to the current algebra relations of Gell-Mann which are not renormalized.  CVC of Feynman and Gell-Mann also hinges upon an analogous situation.\footnote{The absence of ``colored dyons''  \cite{CDyons} mentioned earlier  can also  be interpreted in this manner.}  The results obtained in the BPS limit  (in  the  limit  $v_{2}/v_{1} \to 0$)  are  thus valid at any  finite values of $v_{2}/v_{1}$ \cite{Duality}.  Thus

  \begin{center}
{\tt    The dual group ${\tilde H}$ is the transformation group $H_{C+F}$, seen in the dual magnetic description.  }
 \end{center}

\subsection{Other symmetry breaking patterns} 
 
The  cases such as 
$SO(2N+3) \to   SO(2N+1) \times U(1) $    or   $USp(2N+2) \to  USp(2N) \times U(1)$,  are particularly interesting, as the groups  $SO(2N+1) $ and $USp(2N)$ are interchanged by the GNOW duality.   
 In the first case,  for instance,  the   GNOW  conjecture states that the monopoles  belong to  multiplets  of the dual group  $USp(2N)$.   Although  there are some hints how such GNOW dual monopoles might emerge naturally in the semiclassical approximations \cite{KF},   there is a strong argument (based on ${\cal N}=2$ supersymmetry and global symmetry \cite{CKM,MKY})   as well as 
 clear  evidence \cite{CKM},  against the appearance of these GNOW monopoles as the  light degrees of freedom.  In other words,  even if they might emerge in a  semiclassical approximation, they do not survive quantum effects. 

It is perhaps not a coincidence that the Seiberg duals of ${\cal N}=1$  supersymmetric theories  do not coincide  always  with  GNOW duals.  

The systems  $USp(2N) \to  U(r) \times  U(1)^{N-r}  \to {\bf 1}$   also is known to possess    light non-Abelian monopoles  in the fundamental representation of the dual group   $SU(r)$ \cite{CKM},  which can be nicely understood by our definition of the dual group.

\section{Confinement near  conformal vacua \label{SCF}}  

A particular class of confining vacua, in which confinement and dynamical symmetry breaking are described by non-Abelian magnetic monopoles
{\it  interacting strongly},  are of great interest.  The vacua we are talking about are known as non-Abelian Argyres-Douglas vacua.  
These are found as a particular  case of $r$ vacua, with $r=N_{f}/2$  of  $SU(N)$ SQCD, 
as well as in the massless limit ($m_{i} \to 0$) of all of confining vacua of $SO(N)$ and $USp(2N)$ theories.     Many other examples of vacua with analogous properties can be found in the context of wider class of ${\cal N}=1$ supersymmetric gauge theories \cite{CDSW}. 

Although the details (the global symmetry, the light-degrees of freedom) depend on the model, there is a common feature in this class of systems which makes these particularly interesting.   Because of dynamics and for symmetry requirement the system chooses to produce non-Abelian  (rather than Abelian) magnetic monopoles as the low-energy degrees of freedom, but cannot produce quite as many of them as to make the effective theory   infrared-free.  

As a consequence,  confinement is caused by the condensation of certain monopole composites rather than by the condensation of single monopoles \cite{MKY}. As non-Abelian monopoles carry flavor quantum numbers of the 
original quarks  (this is necessary for the low-energy theory to have the correct symmetry of the underlying theory),  the pattern of the symmetry breaking reflects such a mechanism.   These considerations have been distilled from studies on this class of systems  and   on the problem of understanding non-Abelian monopoles discussed  in various parts of this lecture.

\section{Quantum chromodynamics}

What does all this teach about QCD?  That the Abelian superconductor picture is probably not the correct picture of real-world  QCD  ($SU(3)$) has been already pointed out. In particular, the fact that the deconfinement and chiral restoration transitions occur at exactly the same temperatures in $SU(3)$ lattice measurement, appears to make the assumption 
that Abelian $U(1)^{2}$ monopoles are responsible for  confinement and chiral symmetry breaking, rather awkward (the remark (v) of Section \ref{sec:exactqb}). 
On the other hand,  in ordinary (non-supersymmetric) gauge theories, the ``sign flip'' of the beta function needed to  make 
the non-Abelian monopoles  recognizable infrared (or intermediate-scale)  degrees  of freedom, is  much more difficult to achieve.   If the dual ``magnetic'' group were again $SU(3)$,  the magnetic monopoles of such a theory  (regularized $Z_{3}$ monopoles?)  would probably interact too strongly  and would form composite monopoles ({\it cfr.}  the point (iii) of  Section \ref{sec:exactqb}).    A small number of light  flavors, dressing  these monopoles with flavor quantum numbers,  would not be sufficient.

We might speculate that the dynamics of  QCD   lies   somewhere between.    The dual  theory could be an 
\be       SU(2) \times U(1)   \qquad  {\rm  or}  \qquad U(2)
\ee
theory,   with magnetic monopoles  in ${\underline 2}$ of the $SU(2)$  group and   moreover we expect them to carry   
flavor $SU_{R}(2) \times SU_{L}(2)$  quantum numbers.  We expect them to  interact strongly, but not too much, and it is possible that the  system is close to a nontrivial infrared fixed point, with relatively  nonlocal dyons present at the same time, as in the SCFT effective low-energy  theories of   the  supersymmetric models discussed in the previous subsection.

Let us assume that they are $M_{a}^{i}$, ${\tilde M}_{j}^{b}, $
with the  (dual) color $a,b$ and flavor indices $i, j$, and   carrying opposite $U(1)$ charges. A condensate of the form  
\be   \brc    M_{a}^{i}  \, {\tilde M}_{j}^{a}  \ckt \sim \Lambda^{2}  \, \delta_{j}^{i} \label{monopco}
\ee
might form,  inducing   confinement and   chiral symmetry breaking  $SU_{R}(2) \times SU_{L}(2) \to SU_{V}(2)$
simultaneously.    It could be that the standard quark condensate
\be   \brc    \psi_{L}^{i}  \, {\bar \psi}_{R\,  j}  \ckt \sim    \Lambda^{3}  \, \delta_{j}^{i}
\ee
is  closely related dynamically to or   induced by the monopole condensation,  Eq.~(\ref{monopco}),  for instance  via the  Rubakov effect \cite{Rubakov}. 

It is interesting that in such a picture, there should be a considerable difference between  a theory with quarks in the fundamental representation and a (unrealistic) theory with quarks in the adjoint representation.   The Jackiw-Rebbi  effect works diffrently in
the two cases.  
In the former case the fermion zero modes  give rise to  {\it bosonic}  multiplet  
of degenerate  monopoles, while in the latter case some of the monopoles become fermions.  
In the theory with adjoint quarks, then, there can be considerable difference between the phenomenon 
of confinement and that of chiral symmetry breaking. 
 There is an ample evidence for such a difference   ({\it e.g.},  different transition temperatures)   in lattice gauge theory, as is well known.

 \section{Summary} 

Non-Abelian monopoles are present in the fully quantum mechanical low-energy effective action of many solvable supersymmetric theories.  They behave perfectly as pointlike particles carrying non-Abelian dual magnetic charges.  They play a crucial role in confinement and in dynamical  symmetry breaking in these theories.   There is a natural identification of these excitations within the semiclassical approach, which involves the flavor symmetry in an essential manner.   It is hoped that such an improved  grasp on the nature of non-Abelian monopoles would one day leads to  a better understanding of confinement in QCD.

\section*{Acknowledgments} 

It is a  great pleasure for me to present these notes in honor  of the 65th birthday of my friend  Gabriele Veneziano. 
With his deep understanding of physics, brilliant intuition, elegance of his logics  and inexhaustible fantasy, 
as well as with his  exemplary human quality, he has been a guide to many of us contemporary and younger generations
of theoretical physicists for so many years.   It is not easy to emulate such a high standard, but I present these lecture notes, with the best of my efforts  and  with a deep sense of gratitude to Gabriele.    Finally I wish to thank many friends and collaborators  who contributed at various stages of this investigation.

\appendix

\section{Semiclassical  ``non-Abelian'' monopoles  } \label{sec:General}  

In this appendix we review some general formulae \cite{EW,GNO}.
These degenerate  monopoles appear in  a system with  the gauge symmetry breaking 
\begin{equation}     G   \,\,\,{\stackrel {\langle \phi \rangle
     \ne 0} {\longrightarrow}}     \,\,\, H  
\label {general}   \end{equation}
with a nontrivial  $  \pi_2(G/H)$ and non-Abelian  $H$.  

The normalization of the generators can be chosen \cite{GNO} so that
the metric of the root vector space is\footnote{In the Cartan basis
  the Lie algebra of the group $G$ takes the form
  \begin{equation}   [H_i, H_k]=0, \qquad (i,k=1,2,\ldots, r);  
    \qquad   [H_i,  E_{\alpha}] = \alpha_i \, E_{\alpha}; \qquad
    [E_{\alpha}, E_{-\alpha}]= \alpha^i  \, H_i; 
  \end{equation}
  \begin{equation}   [E_{\alpha}, E_{\beta} ]=  N_{\alpha \beta}\,
    E_{\alpha + \beta} \qquad (\alpha+ \beta \ne 0).  
  \end{equation}
  $\alpha_i =(\alpha_1, \alpha_2,\ldots)$ are the root vectors.  }
\begin{equation}    g_{ij} =  \sum_{roots} \alpha_i \alpha_j  = \delta_{ij}.   \label{rootnormal} \end{equation}
The  Higgs field vacuum expectation value  (VEV)    is  taken to be of the form 
\begin{equation}     \phi_0  =  {\bf h} \cdot  {\bf H},  
\end{equation}
where  $ {\bf h} = (h_1, \ldots, h_{{\textup{rank}(G)}})$ 
 is a constant vector representing 
the VEV.        The root vectors  orthogonal to    $ {\bf h}$  belong to the unbroken  subgroup 
$H$.

The monopole solutions are constructed from various  $SU(2)$  subgroups of $G$ that do not commute with $H$,   
\begin{equation}     S_1=  \frac{ 1}{ \sqrt{ 2 { \bf \alpha}^2}  }  (  E_{{ \bf \alpha}}   +   E_{-{ \bf \alpha}}     ); \qquad  
 S_2=  - \frac{ i }{ \sqrt{ 2 {\bf \alpha}^2  }    }(  E_{{ \bf \alpha}}   -     E_{-{ \bf \alpha}} ); \qquad  
S_3=   {\bf \alpha}^{*} \cdot  {\bf H }, 
\label{su2g}\end{equation}
where ${\bf  \alpha}$    is a root vector  associated with a pair of    {\it broken}    generators    $ E_{\pm{ \bf \alpha}}$.    
   $\alpha^{*}$ 
is a dual root vector    defined by 
\begin{equation}   \alpha^{*} \equiv  \frac{ \alpha }{ \alpha \cdot \alpha}.  
\end{equation}
 The symmetry breaking  (\ref{general})  induces the Higgs mechanism in such an $SU(2)$  subgroup, 
$    SU(2)  \to   U(1).
$
By  embedding the known   't Hooft-Polyakov monopole \cite{TH,BPS}     lying in  this subgroup and 
 adding a constant term to $\phi$   so that it behaves  correctly  asymptotically,  
one easily constructs a  solution of the 
equation of motion \cite{EW,BK}:  
\begin{equation}   A_i({\bf r})  =  A_i^a({\bf r},  {\bf h} \cdot {\bf \alpha}) \, S_a;  \qquad \phi({\bf r}) =   
  \chi^a({\bf r},  {\bf h} \cdot {\bf
\alpha})
\, S_a   +  [ \,  {\bf h}   -   ({\bf h}
\cdot {\bf \alpha}) \,    {
\bf
\alpha}^{*}  ]  
\cdot {\bf H},  
\label{NAmonop}\end{equation}
where 
\begin{equation}    A_i^a({\bf r}) =  \epsilon_{aij}  \frac{ r^j }{ r^2}  A(r); \qquad   \chi^a({\bf r}) =  \frac{ r^a }{ r} \chi(r), \qquad    \chi(\infty)=
  {\bf h} \cdot {\bf \alpha}
\end{equation}
is  the standard   't Hooft-Polyakov-BPS     solution.  Note that $\phi({\bf r}=(0,0,\infty) ) = \phi_0.$  

The mass of a BPS monopole is then  given by 
\begin{equation}      M=\int d{\bf S} \, \cdot    \Tr  \, \phi \,  {\bf B}, \qquad   {\bf B}=  {    r_i  ({\bf S}\cdot \frac{\bf r})  }{ r^4}.  \label{mfield}\end{equation}
This    can be computed  by going to the gauge in which 
\begin{equation}  {\bf B}=  \frac{  {\bf r}    S_3    }{ r^3}  = \frac{  {\bf r}      }{ r^3} \,  {\bf \alpha}^{*} \cdot  {\bf H}, \label{gaugeb}    \end{equation}
to be  
\begin{equation}    M= \frac{ 4 \pi   h_i   \alpha^*_j   }{  g  }     \,   \Tr  \, {H_i\, H_j  }.
\end{equation}
For instance, the mass of the minimal monopole of $SU(N+1) \to SU(N)\times U(1)$  
 can be found easily by using Eqs.(\ref{SUNGEN})-(\ref{formula2})
\begin{equation}     M=  \frac{ 2 \pi  \, v \, (N+1)  }{ g}.
\end{equation}
For the cases $SO(N+2) \to SO(N) \times U(1)$ and $USp(2N+2) \to USp(2N) \times U(1)$, where  
$     \Tr  {H_i\, H_j  } =   C \,   \delta_{ij}, 
$  one finds  
\begin{equation}    M= \frac{ 4 \pi   \,  C \,  {\bf h} \cdot  \alpha^*   }{  g  } =    \frac{4 \,\pi \, v}{g  }, 
\end{equation}
while   for  $SO(2N) \to SU(N) \times U(1), $    $SO(2N+1) \to SU(N) \times U(1), $   and $USp(2N) \to SU(N) \times U(1)$, 
the mass is 
\begin{equation}    M= \frac{ 8 \pi   \,  C \,  {\bf h} \cdot  \alpha^*   }{  g  } =    \frac{  8  \, \pi \, v}{g  }. \label{masssosu} 
\end{equation}

In order to get the $U(1)$ magnetic charge\footnote{In this
  calculation it is necessary to use the generators normalized as $\Tr
  \, T^{(a)} \, T^{(b)} = \frac{1 }{ 2} \delta_{ab},$ such that ${\bf B}=
  {\bf B}^{(0)} \, T^{(0)} + \ldots.  $},   we first divide by an appropriate
normalization factor in the mass formula Eq.(\ref{mfield})
 \begin{equation}      F_m =\int d{\bf S} \, \cdot    \frac{\Tr  \, \phi \,  {\bf B}  }{  N_{\phi}}= \int d{\bf S} \, \cdot  {\bf  B}^{(0)}, \qquad   {\bf B}=  {    r_i  ({\bf S}\cdot \frac{\bf r})  }{
r^4}. \end{equation}
  The result,  which is equal to   ${ 4\pi   g_m}$ by definition, gives the magnetic charge.    The latter  must then be expressed
as a function of the  minimum $U(1)$ electric charge   present in the given theory,   which can be easily   found from the normalized  (such that    $\Tr \, T^{(a)}
\, T^{(a)} = \frac{1}{ 2}$)    form of the relevant    $U(1)$  generator.  

   For example, in the case of the symmetry breaking, $SO(2N) \to  U(N)$,   the adjoint VEV is of the form,  $\phi=  \sqrt{4N} \, v \, T^{(0)},$  where $ T^{(0)}$ is
a $2N \times 2N$  block-diagonal  matrix with $N$    nonzero  submatrices $ \frac{i }{  \sqrt{4N} } 
\left(\begin{array}{cc}0 & 1 \\-1 & 0\end{array}\right). $
Dividing the mass  (\ref{masssosu}) by $\sqrt{N} \, v$  and identifying the flux with $4 \pi g_m$  one gets  $g_m=  \frac{ 2 }{  \sqrt{N} \, g}$.   Finally,  in
terms of the minimum  electric charge of the theory  $e_0=   \frac{ g }{  \sqrt {4N}}$    ( which follows from the normalized form of $T^{(0)}$ above)   one finds 
\begin{equation}   g_m=   \frac{ 2 }{  \sqrt{N} \, g} =    \frac{ 2 }{ N}   \cdot \frac{ 1 }{ 2 \, e_0}.  
\end{equation} 
The calculation is similar in other cases. 

The asymptotic gauge field can be written as 
\begin{equation}   F_{ij} =  \epsilon_{ijk} B_k = 
\epsilon_{ijk}  \frac{ r_k 
}{     r^3}  ({ \beta} \cdot  {\bf H}),    \qquad   { \beta} = {\alpha^*}       \end{equation}
in an appropriate gauge (Eq.(\ref{mfield})).  The Goddard-Nuyts-Olive quantization condition \cite{GNO} 
\begin{equation}  2 \, {\beta \cdot \alpha} \in  { \bf Z} 
\end{equation}
then reduces to the well-known theorem that  for two root vectors  $\alpha_1,\, \alpha_2$    of  any group,
\begin{equation}  \frac{ 2 \,  (\alpha_1  \cdot  \alpha_2) }{ (\alpha_1  \cdot  \alpha_1) } 
\end{equation}
is an integer.     

\section {Root vectors and  weight vectors    \label{sec:Roots}}   
 
\subsection { $A_N=  SU(N+1)$}

It is sometimes convenient to have the root vectors and weight vectors of the Lie algebra $SU(N+1 )$ as vectors in
an $(N+1)$-dimensional space rather than an $N$-dimensional one.  The root vectors are then simply 
\begin{equation}
  (\cdots,   \pm 1, \cdots, \mp 1, \cdots).
\end{equation}
($\cdots$  stand for zero elements)   which all  lie on the plane 
\begin{equation}  x_1+x_2 + \ldots + x_{N+1}=0, \label{sunplane} 
\end{equation}
while the weight vectors are projections in this plane of the orthogonal vectors
\begin{equation}
  \vec{\mu} = (\cdots, \pm1, \cdots)\end{equation}
 where the dots represent zero elements. 

In order to use the general formulas of Weinberg and Goddard-Olive-Nuyts we normalize these vectors so that 
the diagonal  (Cartan)  generators may be written  
\begin{equation}
{\bf{H_i} }  = \diag \,  ( w_{1}^{i}, \,  w_{2}^{i}  \ldots,  w_{N}^{i}, \, w_{N+1}^{i}  \, ),   
 \qquad         i=1,2...N
\label{SUNGEN}  \end{equation} 
where  
   $w_{k}$   represents the $k$-th  weight vector of the fundamental representation 
of $SU(N+1)$, satisfying 
\begin{equation}  {\bf w}_k \cdot    {\bf w}_l =  -  \frac{ 1 }{ 2 (N+1)^2};  \quad (k \ne l); \qquad   {\bf w}_k \cdot  {\bf w}_k   =   \frac{N }{ 2  (N+1)^2},  \qquad
k,l=1,2,\ldots, N+1;
\label{weightsun}\end{equation}
and   $ \sum_{k=1}^{N+1}  {\bf w}_k=0.$  
They are vectors lying  in an $N$-dimensional  space (\ref{sunplane}):  in the coordinates of the
$N+1$-dimensional space,   
\begin{equation}     {\bf w}_i=  \frac{ 1}{ \sqrt{2 (N+1)^3}}  (-1, \ldots, -1, N,-1,-1,\ldots).   \end{equation}
The root vectors are simply
\begin{equation}   \alpha =  {\bf w}_i -  {\bf w}_j  =\frac{ 1}{ \sqrt{2(N+1)} }   (\cdots, \pm 1, \cdots, \mp 1, \cdots) \label{rootssuN}
\end{equation}
with  the norm  \begin{equation}  \alpha \cdot   \alpha = \frac{ 1 }{ N+1}. \label{normroot}  \end{equation}   
Note that for $i \ne j$
\begin{equation}  \Tr \, (H_i \,  H_j) =  w_1^i w_1^j +    \ldots + w_{N+1}^i w_{N+1}^j =  \frac{ -2 N + N-1 }{ {2 (N+1)^3}} =  - \frac{ 1 }{  2 (N+1)^2},
\label{formula1}\end{equation}
while 
\begin{equation}  \Tr \, (H_i \,  H_i) = \frac{N^2 + N}{  {2(N+1)^3}} =   \frac{ N }{ 2 (N+1)^2 }.\label{formula2}\end{equation}

The adjoint VEV  causing the symmetry breaking $SU(N+1) \to  SU(N) \times U(1)$ is of the form, 
\begin{equation}  \phi =   {\bf h} \cdot {\bf H},   \qquad  {\bf h}=   v \sqrt{2 (N+1)^3}   \,  (0,0,\ldots,1). 
\end{equation}

\subsection { $B_N  =   SO(2N+1)$}
The  $N$   generators in the Cartan subalgebra  of the Lie algebra $SO(2N+1)$  can be taken to be 
{\small   \begin{equation} 
   H_i = \left(\begin{array}{ccccc}-i  w_1^i {\mathbf J} &  &  &  &  \\ & - i  w_2^i {\mathbf  J} &  &  &  \\ &  & \ddots &  &  \\ &  &  &  -i  w_N^i  {\mathbf J}   &  \\0 &  &  &  & 0\end{array}\right), \qquad    {\mathbf J}= \left(\begin{array}{cc} & 1 \\-1 & \end{array}\right)   \label{songene2} \end{equation} 
}  where   ${\bf w}_k$ ($k=1,2, \ldots, N$)  are the weight vectors of the fundamental representation, 
which are     vectors   in  an $N$-dimensional Euclidean space
\begin{equation}  {\bf w}_k \cdot    {\bf w}_l =0;  \quad k \ne l; \qquad     {\bf w}_k \cdot  {\bf
w}_k   =   \frac{1  }{ 2(2N-1)}:  \end{equation}
 they form a complete set of orthogonal  vectors. 
The root vectors of  $SO(2N+1)$  group  are   $\alpha=  \{\pm {\bf w}_i, \,\,   \pm {\bf w}_i \pm   {\bf w}_j \}$;
their duals are:
\begin{equation} \alpha^*=   \pm 2(2N-1) \, {\bf w}_i,\qquad   (2N-1) [ \,\pm {\bf w}_i \pm   {\bf w}_j ].
\end{equation}
The diagonal generators satisfy  
\begin{equation}  \Tr \, H_i \, H_j=   \frac{1}{ 2N-1}   \,   \delta_{ij}. 
\label{trhihj}\end{equation}
In the system with symmetry breaking  $SO(2N+1) \to SO(2N-1) \times U(1)$  the adjoint scalar VEV is
\begin{equation}  \phi =   {\bf h}\cdot {\bf H},   \qquad  {\bf h}=  i v \sqrt{2(2N-1)} \,  (0,0,\ldots,1). 
\end{equation}

\subsection{$ C_N   =    USp(2N)$}

The  $N$     generators in the Cartan subalgebra  of $USp(2N)$  are the following         $2N\times 2N$ matrices,   
\begin{equation}  {\bf H }_i=    
\left(\begin{array}{cc} {\bf{B_i} }  & {\bf{ 0 }}  \\ {\bf{ 0 } }  &  - {\bf{B_i}^{t}  } \end{array}\right),      \quad  i=1,2,\ldots,  N, 
\end{equation}
where 
\begin{equation}
{\bf{B_i} }  =
\left( \begin{array}{ccccc}
w_{1}^{i} & & & & \\   
 & w_{2}^{i} & & & \\
 & 0 & \ddots & 0 & \\
 & & & w_{N-1}^{i} & \\  
 & & & & w_{N}^{i}
\end{array} \right) ,   \qquad         i=1,2...N.
\label{USPGEN}   \end{equation} 
The weight vectors  ${\bf w}_k$ ($k=1,2, \ldots, N$)  form a complete set of orthogonal  vectors
  in  an $N$-dimensional Euclidean space and   satisfy 
\begin{equation}  {\bf w}_k \cdot    {\bf w}_l =0;  \quad k \ne l; \qquad     {\bf w}_k \cdot  {\bf w}_k   =   \frac{1  }{ 4(N+1)}.  \label{weighusp}\end{equation}
The root vectors of  $USp(2N)$  group  are   $\alpha=  \, \{  \,\pm  \, 2  \, {\bf w}_i, \,\,   \pm {\bf w}_i \pm   {\bf w}_j \}$.
The diagonal generators satisfy  
\begin{equation}  \Tr \, H_i \, H_j=   \frac{1}{ 2 ( N+  1)}   \,   \delta_{ij}. 
\label{trhihj2}\end{equation}
For the breaking  $USp(2N) \to USp(2(N-1)) \times U(1)$  the adjoint scalar VEV is
\begin{equation}  \phi =   {\bf h}\cdot {\bf H},   \qquad  {\bf h}=   v \sqrt{4(N+1)} \,  (0,0,\ldots,1 ). 
\end{equation}

\subsection{$ D_N =  SO(2N)$}
The  $N$   generators in the Cartan subalgebra  of the $SO(2N)$  group  can be chosen to be 
{\small   \begin{equation} 
   H_i =  \left(\begin{array}{ccccc} -i  w_1^i   \left(\begin{array}{cc}0 & 1 \\-1 & 0\end{array}\right)
 &  &  &  &  \\ &  - i  w_2^i \left(\begin{array}{cc}0 & 1 \\-1 & 0\end{array}\right) &  &  &  \\ &  & \ddots &  &  \\ &  &  &  -i  w_N^i \left(\begin{array}{cc}0 & 1 \\-1 & 0\end{array}\right) &  \\ &  &  &  & \end{array}\right)
\label{songene3} \end{equation}  
} where  ${\bf w}_k$ ($k=1,2, \ldots, N$)  are the weight vectors of the fundamental representation, 
living in  an $N$-dimensional Euclidean space and   satisfying
\begin{equation}  {\bf w}_k \cdot    {\bf w}_l =0;  \quad k \ne l; \qquad     {\bf w}_k \cdot  {\bf w}_k   =   \frac{1  }{ 4 (N-1)}: \label{weighso2}\end{equation}
they form a complete set of orthogonal  vectors. 
The root vectors of  $SO(2N )$  are   $\alpha=  \{   \pm {\bf w}_i \pm   {\bf w}_j \}$.
The diagonal generators satisfy  
\begin{equation}  \Tr \, H_i \, H_j=   \frac{1}{ 2(N-1) }   \,   \delta_{ij}. 
\label{trhihj3}\end{equation}
In the system with symmetry breaking  $SO(2N) \to SO(2N-2) \times U(1)$  the adjoint scalar VEV takes the form
\begin{equation}  \phi =   {\bf h}\cdot {\bf H},   \qquad  {\bf h}=  i v \sqrt{4(N-1)} \,  (0,0,\ldots,1). 
\end{equation}

\section{Seiberg-Witten curves for $SU(2)$   ${\cal N}=2$  super Yang-Mills theory\label{SWcurve}}

The variable $a$ and $a_{D}$  are to be considered as local variables, describing the low energy effective action in a particular patch of the space of vacua (QMS). On the other hand,  the variable $u =\Tr \, \brc \Phi^{2} \ckt $  is a gauge invariant and apparently unique and global variable describing the QMS.    The space  $(a_{D}, a)$  is the covering space ${\tilde {\cal M}}$ of  the space ${\cal M}$ whose coordinate is the complex VEV $u$.   If the base space were simply connected, the map ${\tilde {\cal M}}\to {\cal M}$ would be trivial. In general, a closed loop of the point $u$  in the base space  induces a discrete transformation, called monodromy group,
among the inverse images of the point $u$ in the covering group. 

The fact that the space ${\cal M}$  is nontrivial follows from the  one-loop beta function, 
\[   \tau_{eff} =  d a_{D}/da =  \frac{\theta_{eff}}{ 2\pi} 
+  \frac{4\pi i }{g_{eff}^{2}} \sim  \frac{i}{2\pi}\, \log a+\ldots  \]
so
\[    F(A) \simeq  \frac{i}{2\pi}   A^{2} \, \log  \frac{A^{2}}{\Lambda^{2}}, \qquad
a_{D}=  \frac{dF(a)}{da} \simeq  \frac{i}{2\pi}\,  (  a\, \log a +  \frac{a}{2} ).
\]
The effect of a loop at large $u\sim a^{2}/2$,    $u \to e^{2\pi i}  u$ is $ a \to e^{\pi i}$, so 
\[  a_{D} \to  - a_{D} + 2 a, \quad  a \to - a, 
\]  
or 
\[   \left(\begin{array}{c}a_D \\a\end{array}\right)\to  M_{\infty} \,\left(\begin{array}{c}a_D \\a\end{array}\right), \qquad 
M_{\infty}=\left(\begin{array}{cc}-1 & 2 \\0 & -1\end{array}\right). 
\]
A singularity at $\infty$ in the $u$ space implies the presence of at least one more singularity at finite $u$. As the theory possesses an  invariance under  spontaneously broken discrete  ${\mathbf Z}_{2}$,  under which $u\to -u$, it is natural   to assume a pair of  singularirties at  $u = \pm \Lambda^{2}$.  The key idea of Seiberg and Witten is that these singularities correspond to the points of $u$ where the 't Hooft-Polyakov monopole becomes massless due to  quantum effects. 
Near $u\sim \Lambda^{2}$ then 
\be    a_{D}(u=\Lambda^{2})=0,\qquad   \tau_{D}=- \frac{da}{d a_{D}} \simeq  -\frac{i}{\pi} \log a_{D},  
\label{standard}  \ee
and \[   a_{D}\sim  c_{0}   (u- \Lambda^{2}),  
\]
where Eq.~(\ref{standard}) is  the standard  beta function  of $N=2$ supersymmetric QED. 
Thus a closed loop  in $u$ around the point $\Lambda^{2}$ induces the monodromy transformation
 \[  a\to a - a - 2 a_{D}; \quad   a_{D} \to  a_{D}, \qquad   M_{\Lambda^{2}} = \left(\begin{array}{cc}1 & 0 \\-2 & 1\end{array}\right).  
 \]
The monodromy  transformation around $-\Lambda^{2}$ follows from the consistency condition,  
\[  M_{\Lambda^{2}} \cdot M_{\Lambda^{2}} =  M_{\infty}.  
\]
The map  $a_{D}(u),$ $a(u),$  with the desired properties is precisely the one  given in  Eq.(\ref{torus})-Eq.(\ref{SWsol}).

\section{One-particle representations of ${\cal N}=1$ and  ${\cal N}=2$ supersymmetry algebra \label{susyalgebra}}
  
\begin{description}

\item{(i)}    For a massive ${\cal N}=1$     supersymmetric   particle  states,   one has  ($P^{\mu}= (M,0,0,0)$)
  \be  \{  Q_{\alp} ,   {\bar Q}_{\adt} \} =  \delta_{\alp \adt}   \, 2 M, \qquad   \alp,  \adt =1,2,
\ee
or, by defining       \be  b_{\alp}^{\dagger} = \frac{ 1}{ \sqrt {2M}}  Q_{\alp}, \qquad    b_{\adt}  =  \frac{ 1}{ \sqrt {2M}}   {\bar Q}_{\adt}. 
\ee
These  can be regarded as two pairs of  annihilation and creation operators, $\{ b_{\adt}, b_{\alp}^{\dagger} \} = \delta_{\alp \adt}.$    The complete set of one particle states can then be conctructed by
defining  the vacuum state by  ($i=1,2$)
\be   b_i \, |0\ckt=0;  
\ee
the full set of states are 
\be    |0\ckt, \quad  b_{1}^{\dagger} |0\ckt, \quad   b_{2}^{\dagger} |0\ckt, \quad    b_{1}^{\dagger} b_{2}^{\dagger} |0\ckt,
\label{states}\ee
they form a degenerate supersymmetry multiplet   (two bosons and two fermions).  
For ${\cal N}$ supersymmetry, the same argument shows that the multiplicity of a massive multiplet is 
\be  \sum_{n=0}^{2 {\cal N} }   {2 {\cal N} \choose    n} =2^{2 {\cal N}}.
\ee

\item{(ii)} Massless   \1N  supersymmetric   particle  states:     In this case it is not possible to go to the rest frame but the momentum can be chosen as $P^{\mu}= (p,0,0,p)$. Then
\be  \{  Q_{\alp} ,   {\bar Q}_{\adt} \} = \left(\begin{array}{cc}2 p & 0 \\0 & 0\end{array}\right)_{   \alp  \adt }
\ee
The state $b_{2}^{\dagger} \, |0\ckt$   have a zero norm.   The particle states are given by the positive norm states,   half of (\ref{states}), 
\be   |0\ckt, \quad  b_{1}^{\dagger} |0\ckt.
\ee
The multiplicily of a massless \1N  supersymmetry multiplet is
\be  \sum_{n=0}^{ {\cal N} }   {2 {\cal N} \choose    n} =2^{ {\cal N}}.\ee 

\item{(iii)}    Massive  ${\cal N}=2$    supersymmetric    particle  states with central charges.  
In the rest frame ($P^{\mu}= (M,0,0,0))$ the supersymmetry algebra reduces to
 \be  \{  Q_{\alp}^i ,   {\bar Q}_{\adt}^j  \} =  \delta^{ij} \, \delta_{\alp \adt}   \, 2 M, \qquad   \alp,  \adt =1,2,\quad i,j=1,2,
\ee
   \be   \{     Q_{\alp}^i ,     Q_{\bet }^j \}   = \epsilon_{\alp \bet} \, \epsilon^{ij}  \,( U  + i  V)
\ee 
  Within an  irreducible representation $U$ and $V$ are just numbers (electric and
magnetic charges of these particles). There are three cases:

 \begin{enumerate}
    \item{ $2M<\sqrt{U^2+V^2}$ :} It is not possible to
find a positive-norm   representation of the algebra;
    \item{$2M=\sqrt{U^2+V^2}$ :} 
    A representation exists with multiplicity $2^{ {\cal
N}}=4$ ({short multiplet})  (these are the so-called BPS saturated case); 
    \item{$2M>\sqrt{U^2+V^2}$ :} A representation exists with multiplicity $2^{ 2{\cal
N}}=16$ ({long multiplet}).
\end{enumerate}

 \end{description}

   Proof:   Define \be \frac{Q_1^1}{\sqrt{2M}}=b_1 \qquad
\frac{Q_2^1}{\sqrt{2M}}=b_2 \qquad \frac{Q_1^2}{\sqrt{2M}}=b_3 \qquad \frac{Q_2^2}{2M}=b_4 \ee \be
-\frac{U}{\sqrt{2M}}=u \qquad -\frac{V}{\sqrt{2M}}=v \ee then \be \{b_i,b_j^\dagger \}=\delta_{ij} \qquad \{b_1,b_4
\}=u+iv \qquad \{b_2,b_3 \}=-u-iv \ee \be \{b_1^\dagger,b_4^\dagger \}=u-iv \qquad \{b_2^\dagger,b_3^\dagger \}=-u+iv
\ee 
Now make the change of  variables \be Q_\alp^1\longrightarrow e^{i\gamma} \, Q_\alp^1 \qquad Q_\alp^2\longrightarrow  Q_\alp^2
\ee 
\be b_1\longrightarrow e^{i\gamma} \, b_1 \qquad b_2\longrightarrow  e^{i\gamma} \, b_2 \ee to have $\{b_1,b_4 \}$
real and positive:
 \be \{b_1,b_4 \}=\{b_1^\dagger,b_4^\dagger \}=\alpha=\frac{\sqrt{U^2+V^2}}{2M} \ee
  \be \{b_2,b_3
\}=\{b_2^\dagger,b_3^\dagger \}=-\alpha \ee
In order to see the spectrum, 
 it is convenient to set 
  \be A=b_1
\cos\vartheta+b_4^\dagger \sin\vartheta,  \qquad B=-b_1\sin\vartheta+b_4^\dagger\cos \vartheta.  \ee 
The condition
$\{A,B\}=\{A, {B}^{\dagger }  \}= 0$ yields $\vartheta=\frac{\pi}{4}$:
$A$ and $B$ satisfy disjoint  anticommutators
 \be \{A,B\}=0,  \qquad \{A,A^\dagger \}=1+\alp,  \qquad
\{B,B^\dagger \}=1-\alp.   \ee 
Thus  if $|\alp|  <   1$ there are two creation operators $A^\dagger$, $B^\dagger$;  
while if   $\alp
=\pm1$   $B^\dagger$ (or $A^{\dagger}$)    creates  zero-norm states.       The same passages for $b_2$ and $b_3^\dagger$
lead   to a similar  result.  
The net result is that  particles with mass $M  > \frac{\sqrt{U^2+V^2}}{2}$ come in ``long multiplets'', with multiplicity $2^{ 2\,{\cal N}}=8$,    while the BPS  particles with mass
$M  = \frac{\sqrt{U^2+V^2}}{2}$   come   in ``short multiplets''
of multiplicity  $2^{{\cal N}}=4$.

\end{document}